\documentstyle[emulateapj]{article}   

\slugcomment{To appear in ApJ}   

\def\Epk{E_{\rm pk}}

\def\FE{F_{\rm E}}
\def\NE{N_{\rm E}}
   
\def\E0{E_{\rm 0}}   
\def\P0{\Phi_{\rm 0}}   
\def\Ep0{E_{\rm p0}}

\def\N0{N_{\rm 0}}   
\def\t0{\t_{\rm 0}}

\def\ab{\alpha-\beta}  
\def\a2{\alpha+2}  
\def\b2{\beta+2}  
\newcommand{\ltsima} {$\; \buildrel < \over \sim \;$}   
\newcommand{\gtsima} {$\; \buildrel > \over \sim \;$}   
\newcommand{\lta} {\lower.5ex\hbox{\ltsima}}   
\newcommand{\gta} {\lower.5ex\hbox{\gtsima}}   

\lefthead{Borgonovo \& Ryde}   
\righthead{The HIC in GRB pulses}

\begin{document}   
    
\title{On the Hardness-Intensity Correlation in Gamma-Ray Burst Pulses}   
\author{Luis Borgonovo and Felix Ryde}   
\affil{Stockholm Observatory, SE-133 36 Saltsj\"obaden, Sweden}

\begin{abstract}   
We study the hardness-intensity correlation (HIC) in gamma-ray bursts
(GRBs).  In particular, we analyze the decay phase of pulse structures
in their light curves.  The study comprises a  sample of 82 long
pulses selected from 66 long bursts observed by the Burst And Transient
Source Experiment (BATSE) on the {\it Compton Gamma-Ray Observatory}.
We find that at least $57 \%$ of these pulses have HICs that can be well
described by a power law.  A number of the other cases can still be
explained with the power law model if various limitations of the
observations are taken into account.

The distribution of the power law indices $\gamma$, obtained by
modeling the HIC of pulses from different bursts, is broad with a mean
of $1.9$ and a standard deviation of $0.7$.  We also compare indices
among pulses from the same bursts and find that their distribution is
significantly narrower.  The probability $p$ of a random coincidence
is shown to be very small ($< 2 \times 10^{-5}$).  In most cases,
the indices are equal to within the uncertainties.  These results
demand a physical model to be able to reproduce multiple pulses with
similar characteristics for an individual burst, but with a large
diversity for pulses from an ensemble of bursts.  This is particularly
relevant when comparing the external versus the internal models.
   
In our analysis, we also use a new method for studying the
hardness-intensity correlation, in which the intensity is represented
by the peak value of the $E F_{\rm E}$ spectrum, where $E$ is the
energy and $F_{\rm E}$ is the energy flux spectrum.  We compare it to the
traditional method in which the intensity over a finite energy range is
used instead, which may be an incorrect measure of the bolometric
intensity.  This new method gives stronger correlations and is useful
in the study of various aspects of the HIC. In particular, it produces a
better agreement between indices of different pulses within
the same burst.  Also, we find
that some pulses exhibit a {\it track jump} in their HICs, in which
the correlation jumps between two power laws with the same index. We
discuss the possibility that the {\it track jump}  is caused by
strongly overlapping pulses.  Based on our findings, the constancy of 
the index is proposed to be used as a tool for pulse identification in
overlapping pulses and examples of its application are given.
\end{abstract}   
   

\section{Introduction} \label{intro}   
   
The analysis of the spectral and temporal evolution of gamma-ray
bursts (GRBs) renders us important clues to the underlying processes
giving rise to the phenomenon. The evolution has been studied both
over the entire burst, giving the overall behavior, and over
individual pulse structures (see, for instance, the review by
Ryde 1999a). Pulses are common features in a GRB  
light curve and appear to be the fundamental constituent of it (see,  
e.g., Norris et al.\ 1996, Stern \& Svensson 1996).  
To characterize the spectral evolution, relations  
between quantities describing different aspects  
of the evolution have been reported.  One important correlation is that  
between the {\it hardness} of the spectrum at a certain time and the  
integrated flux up to that time, the fluence; the  
Hardness-Fluence Correlation. 
In the context of these studies, the {\it hardness}
is usually given by a ratio of counts in different energy channels or
by some characteristic spectral energy, such as the peak energy.
Another correlation, which has received  attention, is that between
the hardness of the spectrum and the instantaneous flux (or
intensity); the Hardness-Intensity Correlation (HIC).  Most studies
concerning these correlations examine them in single pulses and do not
compare the behavior of pulses within a burst.  However, this has been
done for the Hardness-Fluence Correlation by Liang \& Kargatis (1996)
and by Crider et~al.\ (1999).  Corresponding studies of the HIC have
been mainly discussed in Kargatis et al.\ (1995).

The main purpose of these studies is to lead us to an understanding of
the emission processes.  The mechanisms that generate the bursts are
still not known.  Many models have been proposed, mostly in the context
of two major scenarios involving relativistic shells.  In the {\em
external} model (M\'{e}sz\'{a}ros \& Rees 1993), a thin shell expands
outward after a single release of energy of unknown origin 
(candidates often considered are cataclysmic stellar collapses or compact
stellar mergers). After an initial $\gamma$-ray quiet phase, the shell
becomes active, perhaps due to interactions with the external medium.
The exact nature of the conversion of the kinetic energy of the  
bulk motion into $\gamma$-rays is
unclear.  In the {\em internal} shock models, a central engine generates a
variable wind and interactions within the wind produce the
$\gamma$-ray emission.  This is often modeled with 
a series of relativistic shells
that are released, with the fast shells catching up with the slow ones,
which leads to the formation of internal shocks.
  
An approach frequently used in these models is to identify each pulse in 
the light curve with a single physical event.  Depending on the  
model chosen, this event could be  the collision between 
inhomogeneities in a relativistic wind in the {\it internal}
models or the ``activation'' of a region on a single {\it 
external} shell.  To validate this reductionistic method, it is 
essential to find common properties among pulses.  

The hardness-fluence correlation was discussed first by Liang \&
Kargatis (1996) who described it as being an exponential decay of the
spectral hardness as a function of the photon fluence. The exponential
decay constant appeared to be invariant between pulses in some
bursts, which led the authors to suggest that the pulses are created
by a regenerative source rather than in a single catastrophic event.
However, Crider et~al.\ (1998a) dismissed the apparent invariance as
coincidental, and consistent with drawing values out of a narrow
statistical distribution, combined with rather large uncertainties in
the determination of the exponential decay constant.

The hardness-intensity correlation was discussed first by Golenetskii
et al.\ (1983, hereafter G83).
In the present work, we study the HIC for
the decay phase of a sample of GRB pulses observed by the Burst and
Transient Source Experiment (BATSE) on the {\it Compton Gamma-Ray
Observatory (CGRO)}. We present an extensive comparison of  
the HIC behavior among pulses from the same burst and  between
bursts, partly motivated by the behavior of the Hardness-Fluence 
Correlation. Some of the results have been presented in preliminary
form in \cite{BR00} and Ryde, Borgonovo, \& Svensson (2000).
  
First, in \S \ref{hic} we discuss previous work and results, in which
the HIC was often found to be a {\it power law} relation. In \S
\ref{fit}, we present the data we used in the analysis and discuss the
observations (\S 3.1), the sample selection (\S 3.2), the spectral
modeling (\S 3.3) and the analysis method used (\S 3.4).  In
particular, in \S 3.4.2, we introduce a new method to analyze the HIC.
We present our results in \S 4. The usefullness of our new analysis
method is shown in \S 4.1, and in \S 4.2 we study individual pulses in
GRBs and present the general distribution of their power law indices
in \S 4.2.1. The cases that were not included in the analysis of the
distribution, are examined in \S 4.2.2. In \S 4.3 we turn to the study
of multi-pulse GRBs and investigate cases with several well separated
pulses (\S 4.3.1) and discuss characteristic cases exhibiting {\it
track jumps} in their HICs. In particular, \S 4.3.2 presents cases
with {\it track jumps} occurring in apparently single pulses.  In \S
4.4 we study the power law HICs of pulses within the same burst and
compare them to the general distribution from \S 4.2.1 and find that
they are more alike than what is expected from the general
distribution.  We discuss the final results of our analysis in \S
\ref{discussion}. Our new analysis method is discussed in \S 5.1 and
the track jumps are interpreted as being the result of heavily
overlapped pulses in \S 5.2.  Finally, in \S \ref{models}, we discuss
how our results impose important constraints on the current physical
models and in particular how they will be relevant in comparing
external versus internal shock models.

\section{The Relation between Hardness and Intensity} \label{hic}

The relation between the hardness and the intensity, during the active
$\gamma$-ray phase of a GRB, has been well investigated. It has been
shown that there is no ubiquitous trend of spectral evolution that can
characterize all bursts; several types of behavior exist.  Norris
et~al.\ (1986) found that the most common trend of spectral evolution
is a hard-to-soft behavior over a pulse, with the hardness decreasing
monotonically as the flux rises and falls.  This conclusion was also
arrived at by Kargatis et~al.\ (1994, hereafter K94).  A few cases exhibited
soft-to-hard and even soft-to-hard-to-soft evolution.  Band (1997)
studied the data from the four, high time-resolution channels from the
Large Area Detectors (LADs) of BATSE. The spectral evolution was
analyzed by auto- and cross-correlating light curves from the different
energy channels.  Most bursts in the sample showed a hard-to-soft
behavior.
    
There are also bursts that do not seem to exhibit any HIC at all, with
an apparently chaotic behavior. The main conclusion drawn by Laros
et~al.\ (1985) and Jourdain (1990), was that there did not exist a HIC
between the spectral evolution and the light curve in their samples
(using PVO and APEX data, respectively).  Over the whole GRB, there
often does not exist any pure correlation, even though the tracks in
the hardness-intensity plane are confined to an area from hard and
bright to soft and weak, indicating an overall trend of increasing
luminosity with hardness (K94).  A seemingly chaotic behavior in that
plane may be the result of a superposition of several short
hard-to-soft pulses that cannot be resolved.  Various types of trends
have also been seen in a single GRB (e.g., Hurley et~al.\ 1992).  The
variety of behaviors are also described in Band et~al.\ (1993) and
Ford et~al.\ (1995).
    
Furthermore, there is another behavior in which the intensity and the
hardness track each other. This behavior is less common than the
hard-to-soft trend and was first noted by G83, 
who described it quantitatively.  They found a power
law relation between the instantaneous energy flux, $F$, and the
energy parameter $E_{\rm 0}$ derived from modeling the photon spectra
using $N_{\rm E}(E) = E^{\alpha} e^{-E/E_{\rm 0}}$, which serves as a
measure of the hardness, i.e.,
\begin{equation}   
F \propto E_{\rm 0}^{\gamma} .   
\label{g83}   
\end{equation}   
The power law index $\gamma$ was found to have a typical value of
1.5--1.7. This value is sometimes referred to in the literature as the
{\it correlation index}.

This analysis was criticized by several workers, including Laros
et~al.\ (1985), Norris et~al.\ (1986), and K94.  It was speculated
that the correlation could possibly be an artifact of the way the
hardness was derived from the two-spectral-channel count rates.
Furthermore, G83 excluded the hard initial phase of the bursts.  Ford
et~al.\ (1995) suggested that the low time-resolution may result in
the initial, hard behavior being missed.  On the other hand, K94
confirmed the description of the HIC made by G83 (eq.~[\ref{g83}]),
i.e., a power law model of the hardness-intensity correlation
(hereafter denoted by PLHIC).  They found a power law correlation in
approximately half of their cases, but with a substantially wider
spread, $\gamma = 2.2 \pm 1.0$.  Furthermore, Strohmayer et~al.\ (1998)
investigated the evolution of the peak energy versus the energy flux
in the {\it Ginga} data and found the PLHIC to be valid here too,
with, for instance, $\gamma \sim 3$ for GRB~890929 (in the 2--400 keV
energy range).
  
The present paper concerns mainly the power-law, hardness-intensity
correlation during the {\it decay phase} of pulses, which is a common
behavior.  Kargatis et~al.\ (1995) found the PLHIC in 28 pulse decays
from 15 bursts, out of a total of 26 GRBs with prominent pulses
studied.
The distribution of the correlation index peaks at 1.7 and has a
substantial spread.  A large spread in the PLHIC index was also found
by Bhat et~al.\ (1994), who studied 19 time structures with short rise
times ($< 4$~s) and slow decays, and concluded that most had a good
correlation between the hardness and the intensity. The value of
$\gamma$ varied from $1.4$ to $3.4$.

Ryde \& Svensson (2000a, 2000b) derived the consequences of combining
the power law model of the HIC and the exponential model of the
Hardness-Fluence Correlation, for the decay phase of GRB pulses.  They
found a self-consistent, quantitative, and compact description for the
temporal evolution of the pulse decay phase.  It was shown that,
assuming the adopted models are valid, the total photon flux must be
$\propto 1/(1+t/\tau)$, where the time $t$ is taken from the start of
the decay and $\tau$ is a time constant that can be expressed in terms
of the parameters of the two empirical correlations.

\section{Data and Methods} \label{fit}   
  
\subsection{Observations}  
    
This work is based on the data taken by BATSE on board the {\it CGRO}
(Fishman et~al.\ 1989). It consists of eight modules placed on each
corner of the satellite, giving full sky coverage. Each module has two
types of detectors: the Large Area Detector (LAD) and the Spectroscopy
Detector (SD). The former has a larger collecting area 
and is suited for spectral continuum studies, while the latter 
was designed for the search of spectral features (lines). For our spectral
analysis we used the
high energy resolution (HER) background and burst data types
from the LADs having 128 energy channels. The burst data have 
a time-resolution of multiples of 64 ms.  The {\it CGRO} Science
Support Center (GROSSC) at Goddard Space Flight Center (GSFC) provides
these data as processed, high-level products in its public archive.
The data are available for all the detectors that triggered on the bursts   
(often 3 or 4 of the detectors closest to the line-of-sight to the   
burst location). Models of the relevant Detector Response Matrix (DRM)   
for each observation are also provided (Pendleton et al. 1995).   
The eight modules of BATSE allow the localization of the GRB,   
needed for the determination of the DRM, since it is dependent on the   
source-to-detector axis angle.
  Finally, we used, for visual inspection of the light curves, the so called
concatenated 64-ms burst data, provided by GROSSC, which is a
concatenation of the three BATSE data types DISCLA, PREB, and DISCSC.

\subsection{Sample Selection}  \label{selection}
    
To construct  a complete sample of strong bursts, we started by selecting
the bursts in the Current BATSE
Catalog{\footnote{http://gammaray.msfc.nasa.gov/batse/
}}, up to GRB~990126 (BATSE trigger number 7353), 
for which it is possible
to measure peak fluxes. These are approximately 80\%
of the totally 2302 observed.  Data gaps and/or missing data types
are the reason why the peak flux for some bursts are not found.
The Current
BATSE Catalog is preliminary, but bursts up to GRB~960825 (trigger~5586) are
published in the 4th BATSE Catalog (Paciesas et al.\ 1999).  The
threshold we chose for accepting a burst was set to a peak flux
(50--300 keV in 1.024 s time resolution) of 2 photons s$^{-1}$
cm$^{-2}$. This selection resulted in a set of 420 bursts.

\begin{figure*}[h]  
\plotone{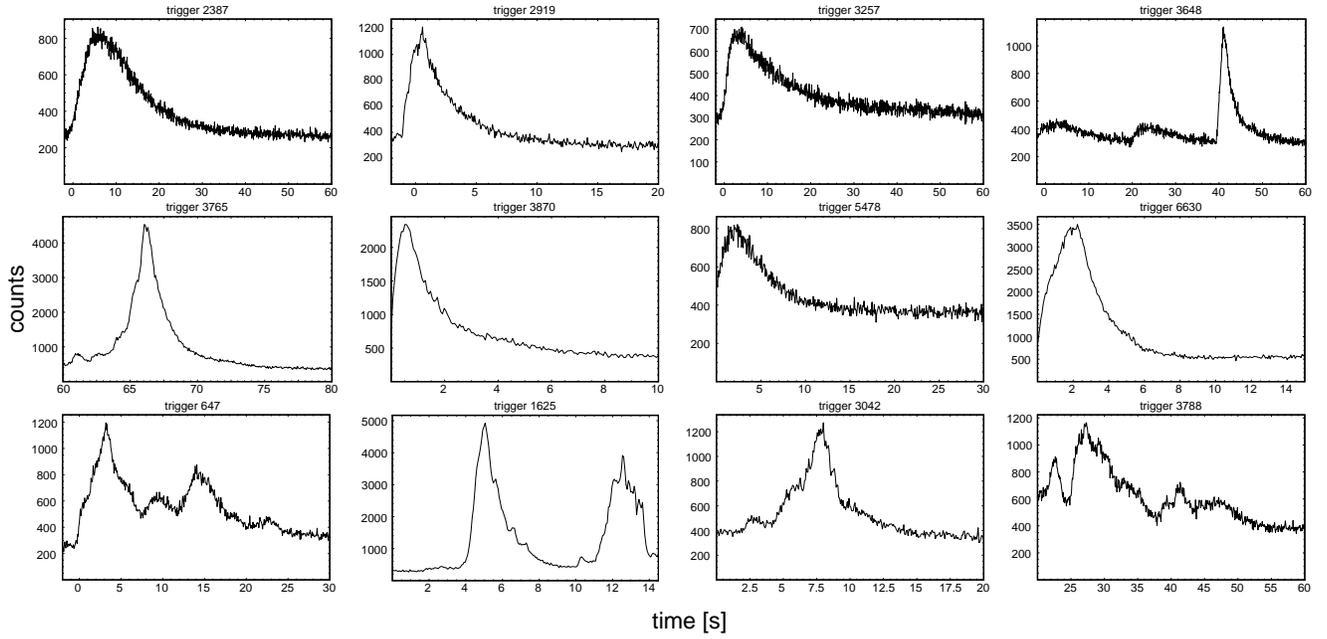}   
\caption{Examples of pulses from our sample. The upper two rows show
fairly smooth, typical pulses selected. However, we also include some
cases that show some apparent sub-structure, as the ones in the bottom
row.  See Table~\ref{Tsample} for further information.}
\label{SampleEx}  
\end{figure*}

This set was examined visually, case by case, using the
concatenated 64 ms time resolution data. We searched for bursts containing
long pulse structures  with a general ``fast rise-slow decay'',
often referred to as ``fast rise-exponential decay'' (FREDs).
No analytical function describing the pulse shape was
assumed. The reason for using such a loose definition is to have a sample
that is independent of any preconceived idea of the pulse shape.
Examples of pulses that are
subsequently included in this broad sample
are shown in Figure~\ref{SampleEx}. 

The identification of the pulse structures may, in some cases, be
disputable. A few algorithms for identification have been introduced
by, e.g., Li \& Fenimore (1996), Pendleton et al.\ (1997), and
Scargle (1998).  Norris et~al.\ (1996) developed a method to
identify pulses based on assuming stretched exponential pulse
shapes. This method was used by, e.g., Crider et~al.\ (1999) to
separate pulses, some of which were  heavily overlapped.  However,
all these methods depend on various assumptions. Our sample
suffers, on the other hand, from some subjectivity, as we
select our pulses visually. However, this should not affect the
results, as these are not strongly dependent on the details in this
selection process.

For the time-resolved spectroscopy, we will use a high signal-to-noise
ratio ($S/N$), which will lead to light curves consisting of only a few
broad time bins.  We need as many time bins as possible to study the
HIC and to arrive at reliable results for the PLHIC index, as we will
be determining its distribution (see \S\ref{single} for details).  For
this purpose we adopt the criterion that the decay phase of the pulses
should have at least 5 time bins with $S/N=30$ to be included in the study.

It should be noted that the first and the last time bins could
partially cover a time interval that is outside the actual HIC behavior of
the decay phase.  For the pulse decay, the first time bin studied
corresponds to the peak of the pulse, and therefore this time interval
might include the
transition between the rise and the decay phase. The
last time bin might also be a transition interval, i.e., the {\it valley}
before the rise phase of the next pulse. 
In the 5-time-bin cases, bearing these qualitative arguments in mind,
there are, in the worst case, only 3 central data points that are 
unaffected and thus are more certain to correctly define the PLHIC index.

The visual inspection resulted in a  
set of 66 bursts, that is $\approx 3\%$ of the original BATSE  
catalog.  These bursts are presented in Table 1 and they constitute our main 
sample, which we have striven to make as unbiased and extensive 
as possible.    
In the Table, the bursts are denoted by both their BATSE catalog and trigger 
numbers.  The detector from which the data were taken and the number of 
time bins ($n_{\rm bins}$) selected for fitting are also presented.  The 
pulses studied are identified by the time $t_{\rm max}$ when their 
count rate is maximal ($t=0$ is the trigger time).  
 
\subsection{Spectral Modeling}  \label{specmodel}
  
The central part of the analysis was performed with the WINGSPAN
package, version 4.4.1 (Preece et~al.\ 1996a).
 The spectral fitting was done
using the MFIT package, version 4.6, running under WINGSPAN.  We
always chose the data taken with the detector which was closest to the 
line-of-sight to the location of the GRB, as it has the
strongest signal (see Table~\ref{Tsample} for the individual cases).
The broadest energy band with useful data was selected, often
25--1900 keV.
A background estimate was made using the HER data, which consist of low 
(16--500~s) time resolution measurements that are stored between 
triggers.  The light curve of the background, during the outburst, was 
modeled by interpolating these data, roughly 1000 s before and after 
the trigger, with a second or third order polynomial fit.

To perform detailed time-resolved spectroscopy it has been shown that
a $S/N \sim 45$ is needed (Preece et~al.\ 1998).
The aim of our spectral analysis, of every time bin, is mainly to
determine the peak energy, $\Epk$, as a measure of the hardness, and allowing
a deconvolution of the count spectrum to find the energy spectrum.
Therefore, we accepted a lower $S/N$,
sometimes down to 30, but we checked that the results are consistent
with higher  $S/N$ ratios. This gives us the possibility to study the burst
pulses with higher time-resolution, which is of great importance for
our study.
     
The background-subtracted photon spectrum, $ \NE(E)$, for each time
bin, is then determined using a forward-folding technique.  An
empirical spectral model is folded through the model of the Detector
Response Matrix and is then fitted by minimizing the $\chi ^2$ (using
the Levenberg-Marquardt algorithm) between the model count spectrum
and the observed count spectrum, giving the best-fit spectral
parameters and the normalization.  The spectra were modeled with the
empirical function (Band et~al.\ 1993):
\begin{equation}   
N_{\rm E}(E) = \left\{ \begin{array}{ll}   
            A \ E^{\alpha}   
e^{-E/E_{\rm 0}} & \mbox{if $(\alpha-\beta) E_{\rm 0} \geq E$}\\   
            A' E^{\beta} & \mbox{if $(\alpha-\beta)   
            E_{\rm 0} < E$ ,} \end{array} \right.    
\label{band}   
\end{equation}   
\noindent   
where $E$ is the energy, $E_{\rm 0}$ is the $e-$folding energy,
$\alpha$ and $\beta$ are the asymptotic power law indices, $A$
the amplitude, and $A'$ has been chosen to make the photon
spectrum $N_{\rm E}(E)$ a continuous and a continuously differentiable
function through the condition
\begin{equation}   
A'=A \left[ (\alpha-\beta)E_{\rm 0} \right]^{\alpha-\beta}   
e^{-(\alpha-\beta)}.
\end{equation}   
The power law indices were always left free to vary if nothing else is 
stated.  The peak energy, $E_{\rm pk}$, at which the $EF_{\rm
E}$-spectrum  ($ \FE = E \NE $) is 
at its  maximum, was used as a measure of the spectral hardness instead 
of $E_{\rm 0}$.  They are related by $E_{\rm pk}= (2+\alpha) E_{\rm 
0}$ and a peak exists only when $\beta < -2 < \alpha$.  The fitting 
procedure has then 4 free parameters: $A$, $\alpha$ and $\beta$, and 
$E_{\rm pk}$.  The photon spectrum arrived at is model-dependent.  
However, as  equation (\ref{band}) often gives 
a good model of the spectra, the photon spectrum found by deconvolving 
the count spectrum should correspond well with the true photon 
spectrum. For every time bin, the instantaneous integrated energy 
flux $F$ was found by integrating  the modeled energy spectrum
over the available energy band of the detector.

These procedures have now given us, beside the spectral parameters, a
data set of peak energies and energy flux values for every time bin of
the pulse decays.  
Note, however, that due to the low energy cut-off of the detector,
in practice only $\Epk$ values $\gtrsim 40$ keV can be reliably
determined. For this reason, it is often the case that a few time bins, at
the end of the pulse decays, are not used in our studies of the
hardness-intensity temporal evolution.

\subsection{HIC Analysis Method}  \label{HICanalysis}

\subsubsection{Statistical Analysis}\label{stat}

Different statistical approaches can be used to test whether a power
law model is adequate to describe the hardness-intensity data, and to
determine its best-fit parameters and their likely uncertainties.
Here, we once again used the minimization of the merit function
$\chi^{2}$.  If the uncertainties of the measurements are known, a
model-independent determination of the goodness-of-fit can be
obtained.  This is done typically by calculating the probability $Q$
of exceeding by chance the $\chi^{2}$ minimum obtained (see, e.g.,
Press et al., 1992).  Data points were weighted using the
uncertainties in the $\Epk(t)$ estimation, which are usually much
larger than the intensity uncertainties.  We found in
too many cases unrealistic values of $Q \approx 1$. 
The most likely explanation is that the uncertainties derived from the
spectral fitting are overestimated. Their usefulness is therefore
doubtful.  One possible cause of this could be the strong correlation
observed between some of the Band et al.\ function parameters,
particularly between $\alpha$ and $A$ (see, e.g., Lloyd \& Petrosian
2000).
Such a correlation could partly be due to an artifact of the fitting.
However, to distinguish this from a physical relation between the 
parameters, a more detailed study has to be undertaken.

The power law model can be linearized by taking the logarithms of the
intensity and the hardness measure.  The power law index then becomes
the slope of the linear relation, and ordinary linear regression can
be used to fit the data.  After this transformation, in most cases,
the $\Epk$-measurement uncertainties are approximately equal and,
since the detector energy channels are approximately equally spaced on
a logarithmic scale, more symmetric error bars are expected. In
addition, the HIC data points are more evenly distributed. Thus,
assuming that the uncertainties are exactly equal, i.e., using the
unweighted method, we can obtain the coefficient of determination,
$R^2$, i.e., the square of the linear correlation coefficient.  For
any given $R^{2}$, $P_{N}(R^{2})$ is the probability that $N$
measurements of two uncorrelated variables would give a coefficient
larger than $R^{2}$. This probability is a measure of how significant
the linear correlation is.

The tests $R^2$, $\chi^2$, and the like, do not have information about
the temporal order of the data points, so obviously one should not infer
anything about the temporal evolution of the HIC from these statistics.
However, most pulse decays in our sample that have a very good PLHIC
also show a good {\it tracking} behavior, i.e., the temporal evolution
in the hardness-intensity plane is monotonic, aside from random
fluctuations that can be attributed to the measurement uncertainties.
Nevertheless, this generalization must be taken with caution in cases
with low correlation coefficients.  Another important aspect not
measured by these tests is whether the residuals show any particular
trend or feature. In this respect, we analyzed in detail the cases
with many time bins and concluded that the data are consistent with
the power law model.

To fit the data directly one can employ a non-linear regression 
numerical algorithm (with the Levenberg-Marquardt method). 
The outcome is essentially the same as in the linear regression and
will not be presented here.

In conclusion, the difference between the indices obtained using  
weighted and non-weighted fittings was, in most cases, within the 
estimated uncertainties.  No significant difference was obtained 
either from our various statistical analyses of the two data sets.  We 
will show the results derived from the non-weighted, linear fittings, 
but we checked that our conclusions are independent of this choice.

\subsubsection{The $\varphi$-method for studying the HIC}\label{phimethod}

The limited spectral coverage of the detector used might affect the
assigned measure of the bolometric flux, i.e., the energy-integrated
flux, as a substantial fraction of the flux could be lost, especially
when the spectrum has a broad shape or peaks close to the boundaries
of the detector. A second problem, when one aims at studying the
correlation within single pulses in the light curve, arises from the
fact that the observed spectra may contain contributions from other
pulses, for instance, previous pulses which still contribute with soft
photons, and unresolved, overlapping pulses.  Furthermore, additional,
separate soft components (Preece et~al.\ 1996b) could also alter the
measured flux value and thus weaken the correlations.  All this can
affect the analysis by changing the shape of the spectrum.

This motivated us to introduce a new representation of the HIC, which
might resolve some of these complications (see also Ryde et al.\
2000).  The value of $E F_{\rm E}$ at $\Epk$ can be used as a
representation of the energy-integrated flux as it, under some
circumstances, is proportional to the total flux.  This quantity
will be denoted by {$\varphi$} (see Fig.~\ref{model}a) and the PLHIC
can be studied as
\begin{equation} 
{\varphi} \propto \Epk ^{\eta}, 
\label{eqeta}
\end{equation} 
where $\eta$ is the new PLHIC index.  This discussion is limited to
the cases where the $E F_{\rm E}$ peak actually exists within the
detector band $[E_{\rm min} , E_{\rm max}]$, which most often is the
case (Band et al.\ 1993).  In the most common case, where 
$E_{\rm max} > (\alpha-\beta) E_{\rm pk} /  (\alpha + 2)$, 
the  proportionality between ${\varphi}$ and $F = \int F_{\rm E} dE$  becomes 
\begin{eqnarray}   
\frac{F}{\varphi} & \equiv & \lefteqn{k(\alpha,\beta,y_{\rm min},y_{\rm max}) =    
\frac{e ^{(\a2)}} {(\a2)^{\a2}} \times} \\ \nonumber   
& & [ \Gamma (\a2) \{ P(\a2,\ab)-P(\a2, y_{\rm min}) \} +  \\  \nonumber 
& & \frac{(\ab)^{\ab} y_{\rm max}^{\b2}-  (\ab)^{\a2} }{   
(\beta+2) e^{\ab}} ] , 
\label{k} 
\end{eqnarray}   
\noindent   
where $y_{\rm min}=(\alpha +2)E_{\rm min}/E_{\rm pk} $, and $y_{\rm
max}=(\alpha +2)E_{\rm max}/E_{\rm pk}$.  $\Gamma(\alpha )$ and
$P(\alpha , y)$ are the gamma function and the incomplete gamma
function, respectively (see, e.g., Press et al.\ 1992).  In the case
that $\alpha$ and $\beta$ depend weakly on $E_{\rm pk}$, the only
$E_{\rm pk}$ dependence in $k(\alpha, \beta, y_{\rm min}, y_{\rm
max})$ is in $y_{\rm min}$ and $y_{\rm max}$.  In particular, when the
flux integration is chosen to be over the whole energy range from $0$
to $\infty$, there will be no dependence at all. Therefore,
$\varphi$ should be a better representation of the
bolometric flux for the study of the HIC.

\section{Results of the Analysis}

\subsection{Correspondence between the $F$ and $\varphi$} \label{Fvsphi}

Before discussing the results of the HIC analysis, we will compare
the HIC relation as given 
by equation (\ref{g83}), but using the parameter $\Epk$ as a measure
of the hardness (\S~\ref{specmodel})
\begin{equation}
F \propto \Epk ^\gamma,
\label{eqgamma}
\end{equation}
and the new representation given by equation (\ref{eqeta}).  For this
comparison, we use a subset of 47 pulse decays (in 39 bursts) from the
sample, which have been selected as having good PLHICs, i.e., the
cases having a relative uncertainty $\Delta\eta/\eta \leq 0.15$ and
marked with an $A$ in Table~\ref{Tsingle} (see \S~\ref{single} for a
discussion of this choice).  For each pulse decay we analyzed the
spectral evolution and determined both the $\gamma$ and the $\eta$
values.
When comparing the coefficients of determination given by the
two  methods
we found that in $83\%$ of the cases $R^{2}_\eta$ is greater than
$R^{2}_\gamma$. To  find whether the 
observed difference is significant, we made use of  Fisher's $z$-transformation
\begin{equation}
z = \frac{1}{2} \ln \left(\frac{1+R}{1-R}\right),
\label{ztrans}
\end{equation}
where each $z$ is approximately normally distributed, with a standard
deviation $\sigma (z) \approx 1/\sqrt{N-3}$, and $N$ being the number
of data points (see, e.g., \cite{Press}). Individual differences are
not significant, but the difference between the respective mean
values, $\bar{z}_\eta - \bar{z}_\gamma$, was found to be positive at a
significance level of $P$-value~$=0.015$ (the $P$-value is the probability 
that the value of the test statistic is as extreme as it is, with the
null hypothesis being true). This implies that
the new $\varphi$-method does give better correlations. 

\begin{figure*}[h]
\epsscale{.3} 
\plotone{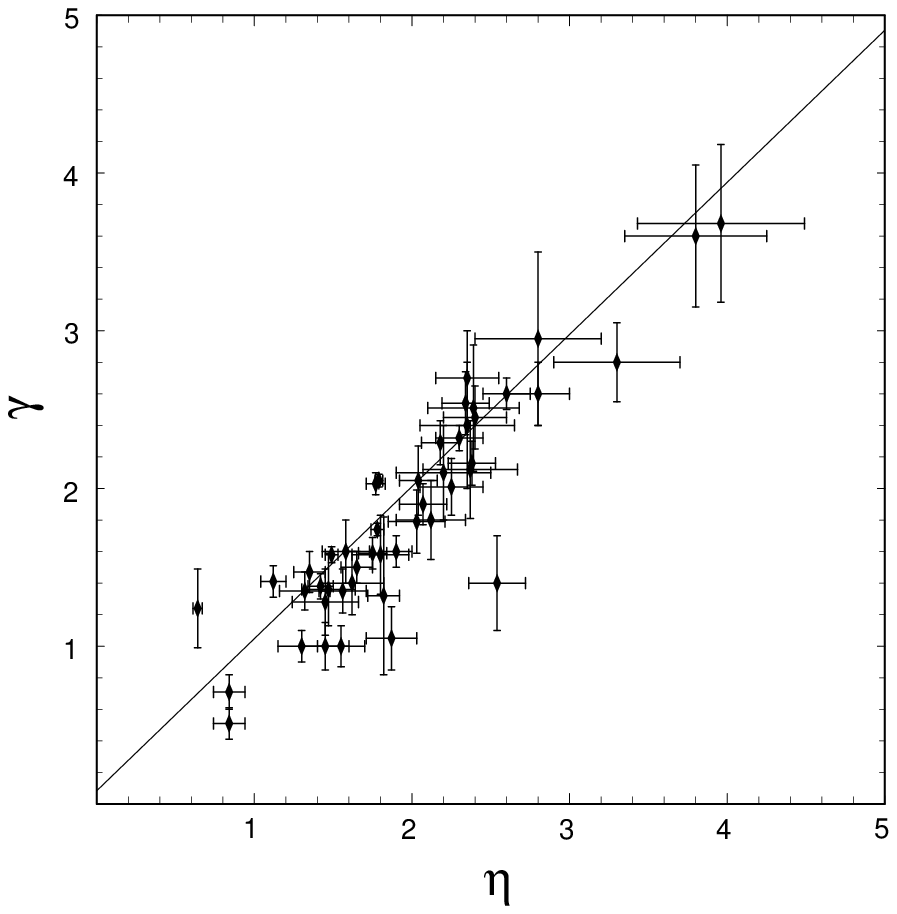} 
\caption{Relation between the correlation indices $\gamma$ and $\eta$ 
for 47 pulses with a good PLHIC, i.e., $\Delta\eta/\eta \leq 0.15$
(marked as subset $A$ in Table~\ref{Tsingle}).  The straight line
shows a weighted linear fit 
$\gamma = (0.96\pm0.09) \eta + (0.08\pm0.16)$ to the data.  This shows
that there is a good correspondence between the two methods, i.e.,
using the intensity measures $F$ and $\varphi$, respectively.}
\label{gammaeta}  
\end{figure*}

The relation between the two PLHIC indices $\gamma$ and $\eta$ for all 
pulse decays in the subset $A$ is shown in Figure~\ref{gammaeta}.  A 
linear fit to these data gives $\gamma = (0.96\pm 0.09) \eta + (0.08\pm0.16)$.  This 
shows that there is a good average correspondence between the two 
methods.  All these results convinced us to use the $\varphi$-method in the
subsequent studies.

\subsection{The PLHIC index of Single Pulses} \label{single}   

The results of the fitting of the bursts in our main sample are
presented in
Table~\ref{Tsingle}.  Apart from the power law indices $\eta$
and $\gamma$ obtained, it shows $R_{\eta}^2$ and its associated probability
$P_{N}(R^{2}_{\eta})$, $R_{\gamma}^2$, and the relative uncertainty $\Delta\eta/\eta$.

From these bursts we  select and study the subset $A$  for which the
hardness-intensity relation is well fitted by a power law model. This
decision is not trivial and the concern was to set a reliable
rejection level to select those cases that are consistent with the
model. The probability $P_{N}(R^{2})$ could be used for
this purpose (see \S \ref{stat}). In practice, the relative uncertainty of
the slope in the linear regression, either $\Delta\eta/\eta$ or
$\Delta\gamma/\gamma$, gives a more convenient measure (at least when the
range of slope values is not close to zero). We found empirically that,
for our set, a rejection level of $P_{N}(R^{2}_{\eta}) < 0.001$
(usually considered as {\it highly significant}) is approximately equivalent
to selecting cases with $\Delta \eta/\eta \leq 0.15$.  This level was chosen
to define the subset $A$ in Table~\ref{Tsingle}. We used the $\eta$ indices
in our selection because the $\varphi$-method gives better correlations
(see \S \ref{Fvsphi}), but almost the same subset would be reached using 
$\Delta\gamma/\gamma \leq 0.2$.
The selection of the rejection level was a trade-off between choosing a
high level and reducing significantly the sample size, or a low one
that would increase the statistics but may allow cases that are likely
also to have large systematic errors.

\subsubsection{The Distribution of PLHIC Indices 
among Bursts} \label{singlePL}

To study the distribution of some parameter associated with GRB pulses,
one should consider that pulses within the same burst may not be
independent. 
Furthermore, they may have a different distribution of the parameter
values considered as compared to their distribution for pulses over
the full burst sample, i.e. the {\it general} distribution.  We will
show below that at least the latter is true for the $\gamma$ and
$\eta$ indices. In that case, a small sample can be easily biased by
taking many values from a multi-pulse burst.  Therefore, to estimate
the general distribution of the PLHIC indices when two or more pulses
were measured in the same burst, only one was taken. To select it, we
chose the one that shows the best correlation in terms of the probability
$P_{N}(R^{2}_{\eta})$.  This method is consistent with the fact that very
often only one pulse, among many in the burst, is found suitable for
this analysis, i.e., we are already disregarding noisy and therefore
poorly correlated cases. On the other hand, since indices are very
similar within bursts, as will be found below, other criteria, 
such as a random selection, produced no significant differences.

We studied how the resulting distributions depend on the chosen
rejection level $\Delta\eta/\eta$.  When trying to estimate the underlying
probability distribution of any sample, care must be taken that the
measurement uncertainties do not contribute significantly. Allowing cases
with larger relative uncertainties, i.e., worse PLHIC cases, results in
almost identical mean values but larger standard deviations. This is
to be expected, since the data have now a larger intrinsic dispersion
that is added to (convolved with) the real distribution. For later use
in our studies, it is important to set a fairly high rejection level, in order
to have a good estimate of the real dispersion.

Thus, taking from the main sample one pulse per burst and only the
cases with $\Delta\eta/\eta \leq 0.15$, defines the subset of 39
pulses marked with $B$ in Table \ref{Tsingle}, for which we will study
the general distribution of the indices (note that $B \subset
A$). Using the $F$-method, the mean value of the PLHIC index is
$\bar{\gamma} = 1.9 \pm 0.1$ with a standard deviation of the sample
$\sigma_{\gamma} = 0.72 \pm 0.08$.  We also provide the expected
deviations of these estimators to facilitate the comparisons.  Using
the $\varphi$-method the mean value and dispersion for the PLHIC
indices are $\bar{\eta} = 2.0 \pm 0.1$ and $\sigma_{\eta} = 0.68 \pm
0.08 $.  The weighted fittings mentioned in (\S~\ref{stat}) produce
very similar results. For example, the values obtained with that
method are $\bar{\eta}_{\rm w} = 2.15 \pm 0.10$ and 
$\sigma_{\eta_{\rm w}} = 0.80 \pm 0.09$. The differences are 
of the same order as the estimated uncertainties.

These results should only be compared with the ones obtained in
studies of the decay part of pulses. As mentioned above, Kargatis
et~al.\ (1995) found a central value of $\bar{\gamma} \sim 1.7$, which
is somewhat lower than ours. However, their sample of 28 pulses was taken
from 15 multi-pulse GRBs. This might produce a slower convergence to
the real mean value. In fact, if we allow more than one pulse per
burst in our statistics, we get a $\bar{\gamma} = 1.8 \pm 0.1$ for 47
pulses from 39 bursts (subset $A$ in Table~\ref{Tsingle}). 
Note that although in many bursts only one pulse was found suitable
for our studies, it does not mean that there were no other pulses
present. A real distinction between single and multi-pulse bursts
would be difficcult to achieve, especially as we do not know what a
single pulse really looks like.

\begin{figure*}[ht] 
\epsscale{0.5}
\plotone{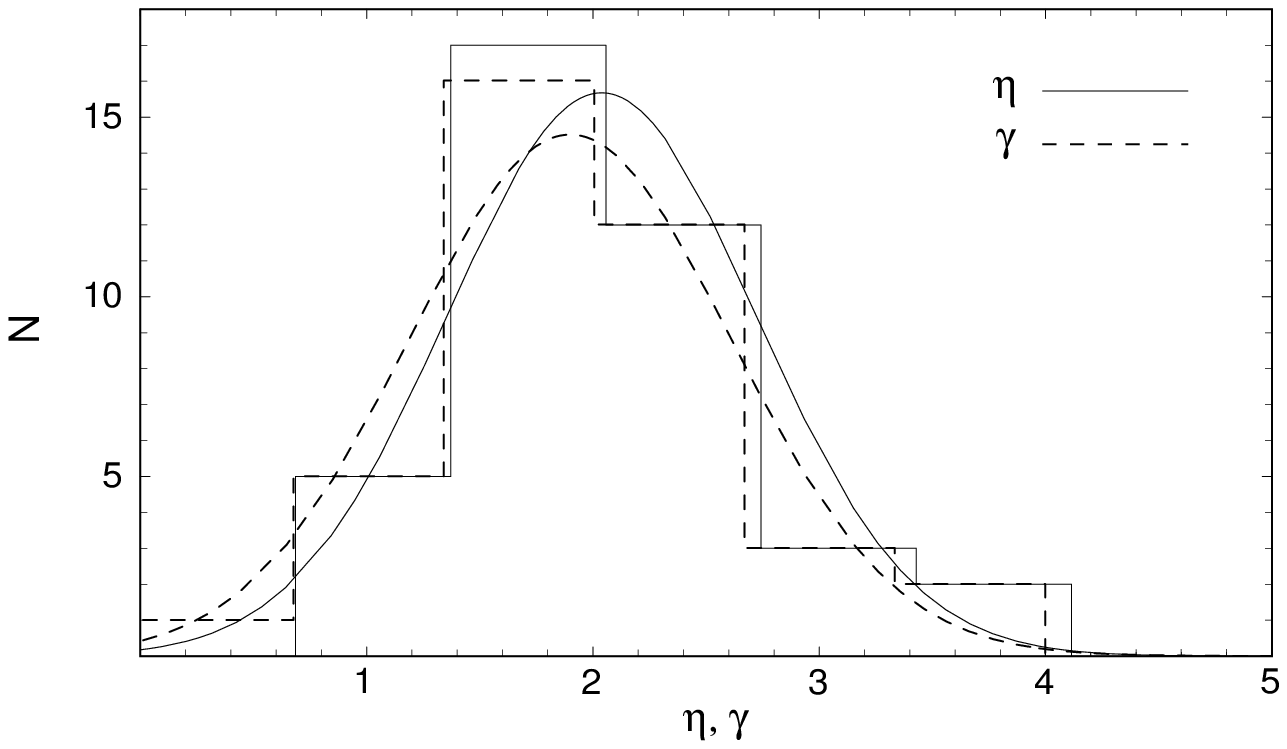}   
\caption{Distributions of the hardness-intensity
correlation indices, $\gamma$ and $\eta$, for a set of 39 pulses
(marked in Table~\ref{Tsingle} as subset $B$). This set is defined by
all the sample cases with $\Delta\eta/\eta \leq 0.15$, but restricted
to only one pulse per burst (see text for more details). Both index
distributions can be approximated by normal distributions with
$\bar{\gamma}=1.9 \pm 0.1$ and $\sigma_{\gamma}=0.72 \pm 0.08$, and
$\bar{\eta}=2.0 \pm 0.1$ and $\sigma_{\eta}=0.68 \pm 0.08$
respectively.}
\label{histo} 
\end{figure*}

In Figure~\ref{histo}, histograms of the subset $B$ of measured PLHIC
indices, $\gamma$ and $\eta$, are displayed. The sample is too small
for a detailed study of the underlying distributions. However, for
later use in this work (\S~\ref{tests}), we test the assumption of a
normal distribution for both indices. We use the Geary test of
normality (see, for instance, Devore 1982), a simpler but more
powerful test procedure than the $\chi^{2}$ test, that is specific for
this purpose. It is based on the ratio of the average absolute
deviation to the square root of the average square deviation, given by
\begin{equation}   
U=\frac{\sqrt{\pi/2} \sum_{i} |X_{i}-\bar{X}|/ n}   
       {\sqrt{\sum_{i} (X_{i}-\bar{X})^{2}/n}} .   
\end{equation}   
\noindent   
For samples larger than 20 the Geary test can be approximated by a
two-tailed $Z$-test, where the null hypothesis  is that the
underlying distribution is normal and the standardized $U$ is
\begin{equation}   
Z=\frac{U-1}{0.2661/\sqrt{n}} .   
\end{equation}   
\noindent   
For the $\eta$ and $\gamma$ samples, high $P$-values are found, 0.73
and 0.34 respectively, implying that the approximation of a normal
distribution can be reliably used.

\subsubsection{The Excluded Cases} \label{badHIC}

We found, using the $R^2$ statistics, that $57 \%$ of the main sample
of pulse decays (subset $A$) could be described by a power law at a
highly significant confidence level. At the level of significance of
only $P_{N}(R^{2}_{\eta}) \leq 0.01$ (approximately equivalent results would be
obtained using $\Delta\eta/\eta < 0.2$) the number increases to about
$75\%$. Since the $R^2$ coefficient is not a very sensitive test,
these percentages should be taken as approximative.

There are several possible causes for some of the cases showing poor
power law correlations. 
We could  often recognize problems that were affecting our
analysis of the HIC in a systematic way, making uncertain any
conclusion  about the
applicability of the power law, or any other model for that matter. In
some cases, for instance, the $EF_{E}$-spectrum peaks at higher
energies than the observed band. However, the numerical algorithm finds
an $\Epk$ value within the energy window using $\beta > -2$. As the
whole spectrum evolves towards lower energies, at some point the
absolute maximum enters the observation window but the overall
evolution of the HIC is poorly fitted by a power law. 

Although the Band et al.\ function proved to be a very flexible model
to describe most GRB spectra, it presents difficulties when trying to
fit some particularly broad shaped cases (see, e.g., Ryde 1999b). In a
graphical $EF_{E}$ vs.\ $E$ representation, a typical zig-zag
evolution is obtained, accompanied by opposite changes in $\alpha$ and
$\beta$. A similar effect occurs when either of the latter two
parameters is loosely constrained by the data. Although in these cases
an overall {\it track} behavior is apparent, PLHIC fits result in poor
$R^{2}$ coefficients, and residuals show an uneven dispersion.

Furthermore, the selected pulse decay could also turn out not to  be
a single pulse decay after all, even if it appears to be. 
Overlapping, ``hidden'' secondary pulses in the light curve produce
a characteristic effect in the HIC evolution that, 
under some circumstances,  can be recognized and used as a pulse 
identification aid. This will be discussed in detail in \S~\ref{jumps}. 
Finally, some cases could actually have another  
HIC behavior. An investigation of alternative models 
will be presented elsewhere (Ryde \& Svensson 2000c, in preparation).

\subsection{The PLHIC Index of Multi-pulse Bursts} \label{multipulse}

We will now study the distribution of the PLHIC indices within a
single burst.  In order to increase the sample, we look for
additional pulses, with 4 time bins, in the GRBs listed in
Table~\ref{Tsample}.  The expanded sample of multi-pulse bursts
includes 4 of these shorter pulses, but in each case there is a longer
pulse to compare with.  The pulses from GRB 950104 (trigger 3345) are also
included, based on the  discussion in \S~\ref{jumps}. In this
way, we increase the number of multi-pulse cases from 11 to 15 GRBs.
They all show at least moderately good correlations.

Apart from these cases, a set of pulses, apparently single, but having
a {\it track jump} in the HIC will be studied in detail. They are
noted with {\it TJ} in Table~\ref{Tsample} and can be
interpreted as consisting of several pulses
(see also Borgonovo \& Ryde 2000).
  
\subsubsection{Bursts with Several Well-separated Pulses}\label{jumps}  
  
\begin{figure}[h!] 
\epsscale{.8}
\plotone{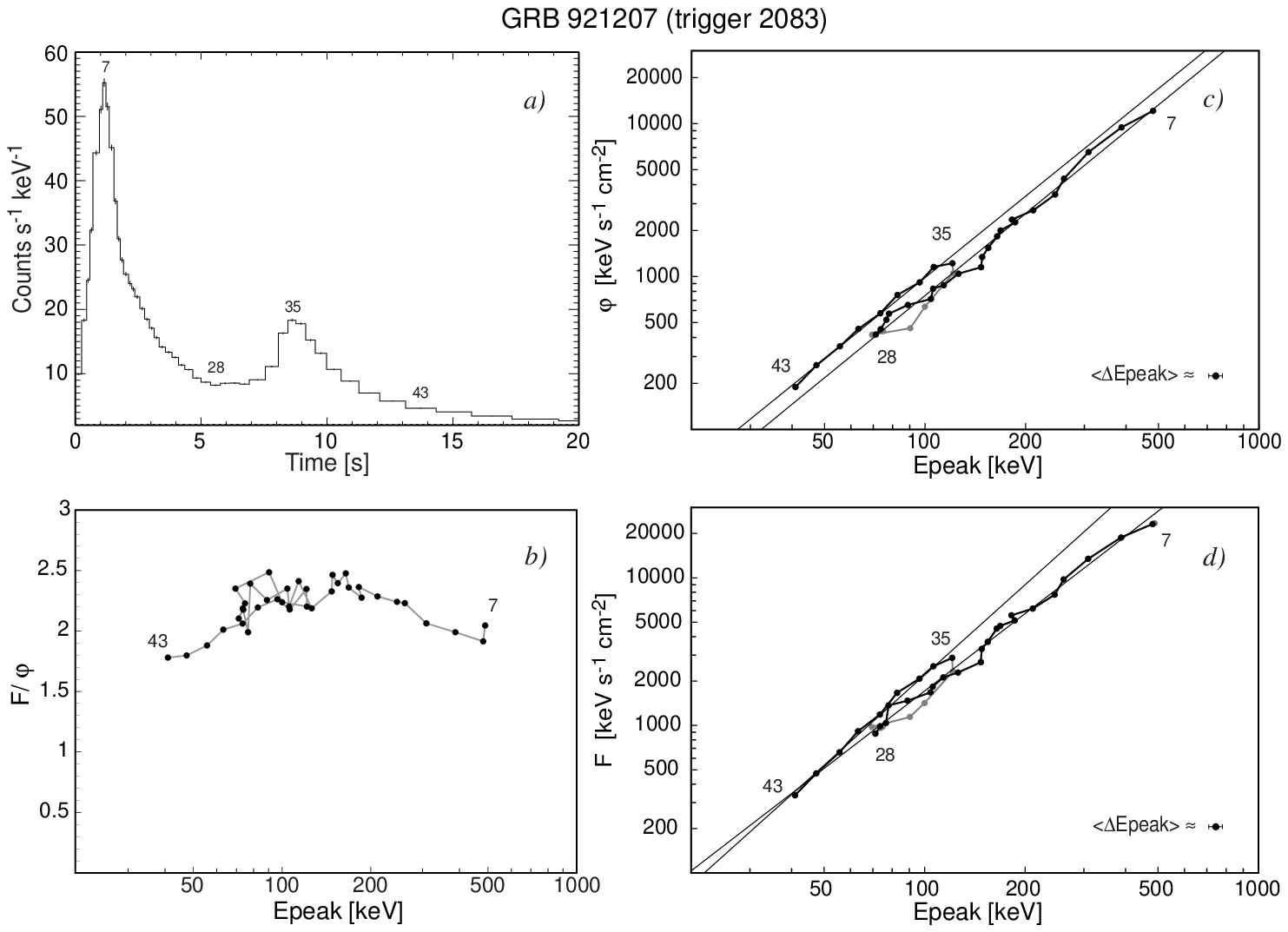}
\caption{$a$) Count rate time history showing the time binning used to
study the spectral evolution of GRB~921207 (trigger~2083). A few bin
numbers are shown. The last few bins are not included in this study
because the $\Epk$ estimation is very uncertain as the peak of the spectrum
approaches the low energy limit of the detector. This is often seen when
$\Epk \lesssim 40$ keV. $b$) Time evolution of the ratio $F/\varphi$. The
integration of the flux was made over the energy range 24--2000
keV. The $F$-measure will suffer from window problems at high and low
energies.  $c$) The PLHIC measured with the $\varphi$-method.  The
evolution during the decay phases is shown with dark dots. Error bars
are excluded for clarity, and only a typical average value is shown
separately. The numbers refer to the time-bins in panel ($a$). We find, for
each pulse decay, the power law indices $\eta_{1}=1.78\pm0.04$ and
$\eta_{2}=1.77\pm0.06$.  $d$) Same as ($c$), now using the
$F$-method. The corresponding indices are $\gamma_{1}=1.74\pm0.04$ and
$\gamma_{2}=2.03\pm0.07$. Note that, as in general, the similarity
between the PLHIC indices of pulses within a burst is more evident
using the $\varphi$-method.}
\label{2083} 
\end{figure}

The sample of GRBs which have two or more well-separated pulses is 
shown in Table~\ref{Tmultiple}.  As an example, we present the 
case of GRB~921207 (trigger~2083) in Figure~\ref{2083}.  This burst 
has been examined by many workers (e.g.,~\cite{Ford95}, Crider et al.\ 
\cite{criderHunta}, \cite{cr99}, and Ryde \& Svensson \cite{RS99a}) 
and is one of the best BATSE cases available for this type of study, 
with two bright and long pulses.  In Figures~\ref{2083}$c$ and $d$, the 
hardness-intensity evolution is shown.  The decay phases of both 
pulses show very good correlations and the power law indices are equal 
to within the estimated uncertainties.
>From the point of view of the PLHIC, the first pulse structure is
consistent with being a single pulse. This is also the conclusion of
Ryde \& Svensson (1999) who  studied various aspects of the
spectral/temporal evolution of GRB pulses. However, with the
pulse identification method proposed by Norris et al.\ (1996) this
pulse structure can be modeled as a superposition of two individual,
overlapping, stretched exponential pulses 
(see Crider et~al.\  1999). In the present
spectral analysis there is no  indication of this, although one cannot 
rule it out.

\begin{figure}[ht!] 
\epsscale{.8}
\plotone{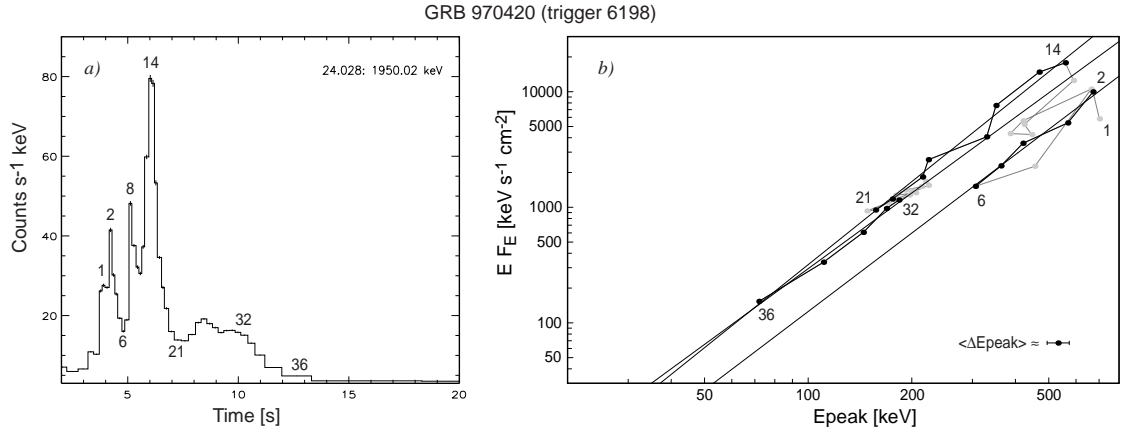} 
\caption{$a$) Count time history of GRB~970420 (trigger~6198)
showing the binning used to study the spectral evolution.  The second 
pulse (at bin number 8) actually consists of two close-lying pulses, which 
are seen in the 64-ms data. $b$) The peaks of the instantaneous 
spectra at successive times are represented by dots. Error bars are 
excluded for better visibility, and only a typical average value is 
shown separately. The numbers refer to the time-bins in panel ($a$). During the 
decay phases (dark dots), the HIC follows very good power laws with indices 
$\eta_{1}=2.25\pm0.20$, $\eta_{2}=2.40\pm0.15$, and 
$\eta_{3}=2.20\pm0.15$, equal to within the estimated uncertainties.} 
\label{6198} 
\end{figure}

As another example, the overall evolution of GRB~970420 (trigger~6198)
is shown in Figure~\ref{6198}. While the spectral evolution during the rise 
phases does not seem to follow any clear trend, again in this case we 
observe a good PLHIC during the decay phases. The slopes are equal 
to within the estimated uncertainties (see Table~\ref{Tsingle}).

\begin{figure}[t!] 
\epsscale{0.7} 
\plotone{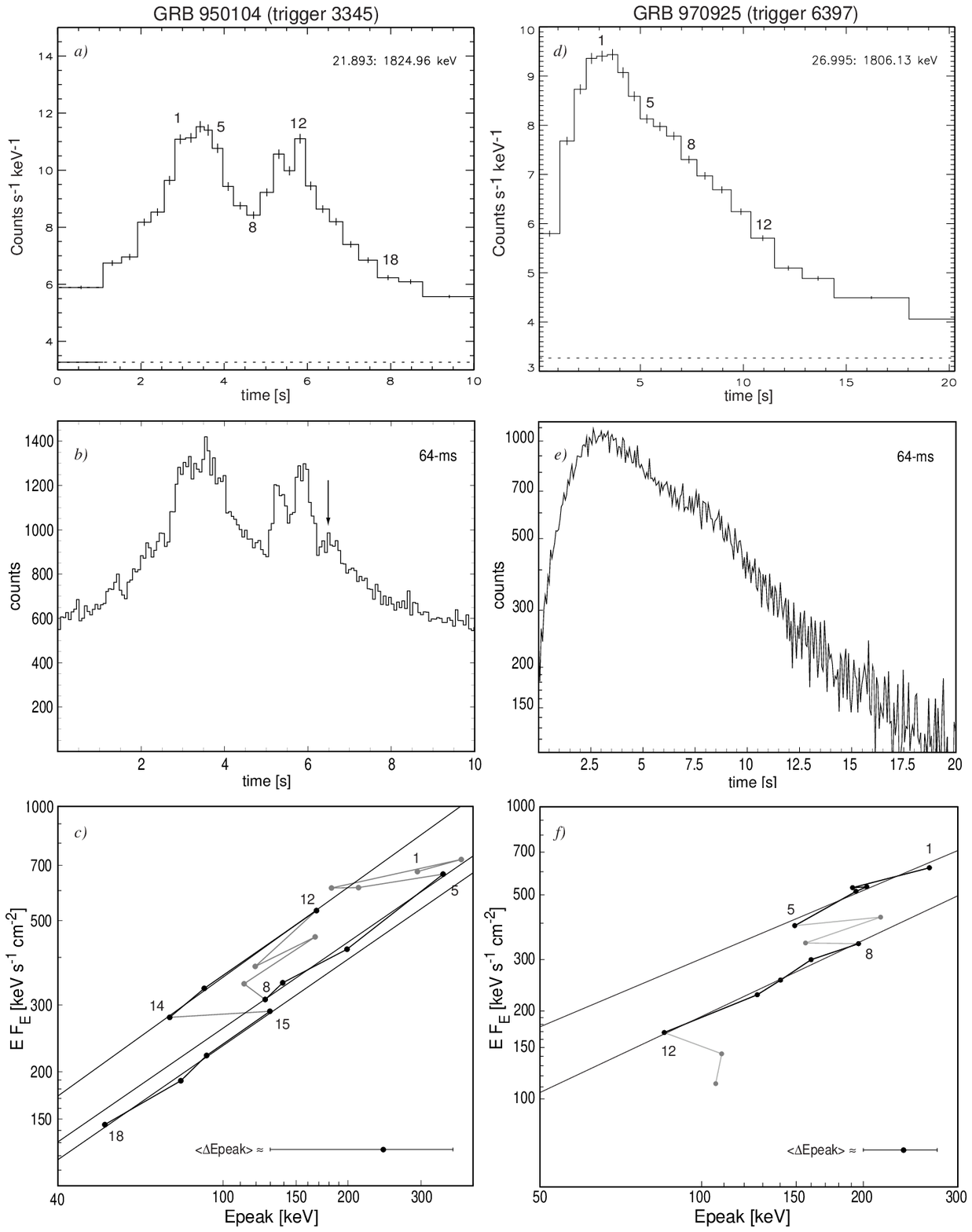} 
\caption{Spectral evolution of GRB~950104 (trigger~3345) and GRB~970925 
(trigger~6397).  $a$), $d$) Photon time history showing the binning used to 
study the spectral evolution ($S/N=40$).  The dashed line indicates 
the background level.  $b$) The same time interval using 64-ms time 
resolution data.  A secondary peak is present at $t \simeq 6.5$s.  See 
the text for details.  $c$) The peaks of the instantaneous spectra at 
successive times are represented by dots.  The numbers refer to the 
time-bins.  The first pulse decay follows a power law with index 
$\eta_{1}= 0.77\pm0.06$.  The second pulse decay initially follows the 
same law, within the measurement uncertainties, with $\eta_{2}= 
0.78\pm0.025$.  A {\it track jump} occurs at the time of the rise of 
the secondary peak in the light curve and then the spectral peak 
evolution continues on a parallel track $\eta_{3}= 0.81\pm0.05$ while 
the light curve decays again.  Error bars are excluded for better 
visibility, and only a typical average value is shown separately.  The 
fitting procedure largely overestimates the uncertainties in these 
cases.  $e$) The burst trigger~6397 at 64-ms time resolution.  $f$) 
Evolution of the peak energy during the pulse decay phase.  As in the 
previous case ($c$), the whole decay presents a poor hardness-intensity 
correlation, but it can be divided into parallel tracks with very good 
HICs and equal indices to within the uncertainties ($\eta_{1}= 
0.80\pm0.15$ and $\eta_{2}= 0.90\pm0.07$).  Note the presence of a 
small ``bump'' at $t \simeq 7.5$ s, i.e., at bins number 6 and 7.}
\label{3345&6397} 
\end{figure}

In Figure~\ref{3345&6397} the light curve of GRB~950104 (trigger~3345) 
is shown. The section after the peak at $t\simeq6$~s
can be interpreted as having two pulses which  overlap closely.
An indication of the existence of a secondary pulse can be found as 
follows. The assumption has to be made that the rise and
decay phases of individual pulses are strictly monotonic and then we calculate
whether the deviation can be attributed to the Poisson noise in the
signal. For that purpose, a simple test would be to find the significance
of the count difference between the suspected peak and the nearby
valleys (see Li \& Fenimore 1996): 
\begin{equation}
C_{\rm peak}-C_{\rm valley} \geq N_{\sigma} \sqrt{C_{\rm peak}} ,
\end{equation}   
\noindent 
where $N_{\sigma}$ is the number of $\sigma$ standard deviations.  In 
the case of GRB~950104, we found a significance 
$N_{\sigma}=3.2$, within the interval $3 \leq N_{\sigma} \leq 5$ recommended 
for the test, i.e., marginally significant.  Figures~\ref{3345&6397}
$a$ and $b$
show two count time history plots of this burst; one  with a time binning 
with  $S/N=40$ and the other with the 64-ms  time 
resolution  of the concatenated data.  The spectral evolution of the 
peak energy corresponding to the time binning in panel ($a$) is also
shown in panel ($c$).  
Initially, the second decay has approximately the same PLHIC index as
the first one.  It then makes a change into a parallel track.
Checking the 64 ms count rate time history, one can see that this {\it
track jump} coincides with the presence of the small secondary peak
(marked with an arrow) which is hidden in the low time resolution
binning.  Even though the significance of the peak is somewhat
marginal, the agreement with the slope of the first pulse track makes
the case more compelling for the presence of overlapping pulses.  The
power law indices obtained for the tracks are $\eta_{1}=0.77\pm0.05$,
$\eta_{2}=0.79\pm0.03$, and $\eta_{3}=0.76\pm0.05$, equal to within
the estimated uncertainties.

Note that the $\Epk$ uncertainties derived from the spectral fittings 
(Fig.~\ref{3345&6397}c) are heavily overestimated. This is a common 
characteristic of the {\it track jump} cases, and it will be discussed in 
\S \ref{phidiscussion}.
  
\subsubsection{Apparently Single Pulses} \label{jumpsII}
  
\begin{figure*}[h] 
\epsscale{0.8}  
\plotone{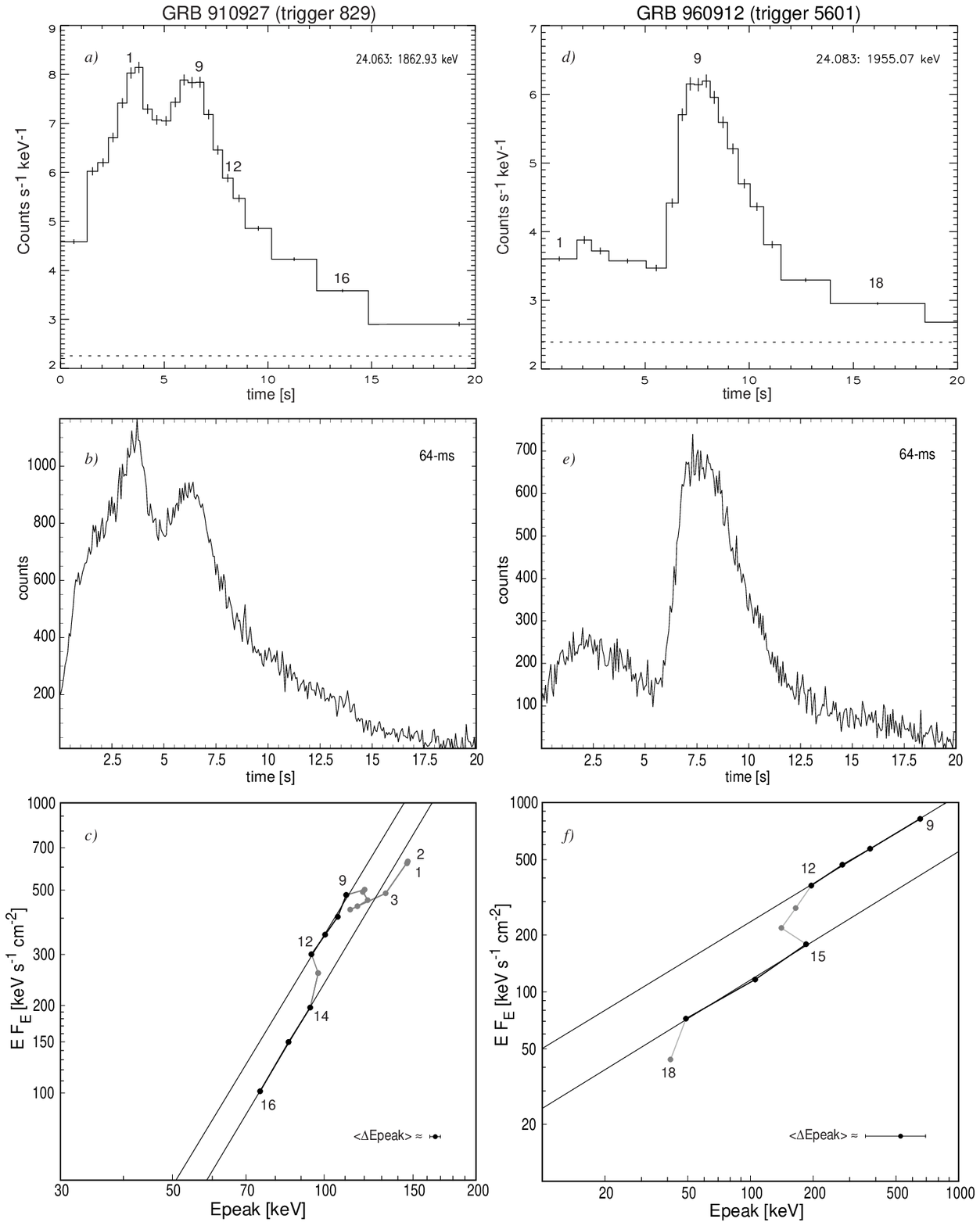}   
\caption{Spectral evolution of GRB 910927 (trigger~829) and GRB 960912 
(trigger~5601).  $a$), $d$) Photon time history showing the binning used to
study the spectral evolution with signal-to-noise levels of 40 and 30,
respectively. The dashed line indicates the background level.  $b$), 
$e$) Time history at 64-ms time resolution.  $c$) Evolution of the peak
energy during the pulse decay phase. As in the previous examples,
it can be divided into two parallel tracks with indices $\eta_{1}=
2.95\pm0.3$ and $\eta_{2}= 2.91\pm0.07$. Note that the first pulse,
that has its peak at bin 2, seems to follow the track defined by the
last three data points numbered 14--16. It could therefore be that the
pulse peaking at bin 9 has a fast decline and  is amid the decay of
a dominating, long pulse. $f$) The same for burst trigger~5601, with
indices $\eta_{1}= 0.675\pm0.01$ and $\eta_{2}= 0.675\pm0.04$, equal
to within the uncertainties. In both cases the high-resolution light
curve shows no clear sign of a ``hidden'' pulse.}
\label{829&5601}  
\end{figure*}

The three cases presented here seem to be single pulses as seen from
the 64 ms light curves.  These pulses are taken from GRB~910927
(trigger~829), GRB~960912 (trigger~5601), and GRB~970925
(trigger~6397), respectively and they are shown in
Figures~\ref{3345&6397} and \ref{829&5601}.  All three behave
similarly to the case GRB~950104 discussed above.  We have modeled
their decay evolution with two power laws and a {\it track jump}. The
result of the fittings are shown in Table~\ref{Tjumps}.

Although there is no clear indication of secondary peaks, in the case
of GRB~970925, a small ``bump'' in the light curve (more easily
noticed in a log-linear graph) coincides with the {\it track jump}.
The measured indices, $\eta_{1}=0.77\pm0.15$ and
$\eta_{2}=0.86\pm0.05$, are again equal to within the uncertainties, as in
the other two cases.

In the case of the pulse decay in GRB~910927, and only in this case,
the fitted results are obtained with the parameter $\beta$ frozen for
all points.  Although the qualitative behavior is the same, it
produces a sharper transition between the tracks.  
Note that the spectral evolution (Fig.~\ref{829&5601}c) during the decay
of the first pulse, given by the bins 2--4, seems to follow the same track
as do the points numbered 14--16. It could therefore be that the last
points, in fact, belong to the decay of the first pulse and that the pulse
peaking at bin 9 has a fast decline and is superimposed in the middle of the
dominating, long first pulse.

\subsection{The Distribution of the PLHIC Index within GRBs}   \label{tests}

An interesting question is how the distribution of the HIC index within 
a burst compares to the general distribution 
as presented in \S \ref{single}.
Our set of multi-pulse bursts was introduced in \S \ref{multipulse}
and listed in Table~\ref{Tmultiple}. It
consists of 35 pulses taken from 15 GRBs. We can see, from
Tables~\ref{Tsingle} and \ref{Tmultiple}, that the PLHIC
indices are, in most cases, very similar.  To estimate the probability
of having similar indices within bursts purely by chance, it will be
assumed first that the set of random variables $X_{i} \in \{\eta\}$
(or $\{\gamma\}$) have approximately a normal distribution 
$X_{i} \sim  N(\bar{\eta},\sigma_{\eta})$, as has been justified in
section \S \ref{singlePL}. Let $Y \equiv X_{i}-X_{j}$ be the difference 
between two indices from the same burst. If they are independent, then 
$Y \sim N(0,\sqrt{2} \sigma_{\eta})$. A hypothesis test can be performed
to see whether $\sigma^{2}_{0} \equiv 2 \sigma^2_{\eta}$ is the true
variance. Thus, choosing the null hypothesis 
$H_{\rm 0}\!:\sigma^{2}=\sigma^{2}_{0}$, we will use first the test statistic
\begin{equation}   
\chi^{2}_{a} = \frac{(n-1) s^{2}}{\sigma^{2}_{0}} ,   
\label{chi2}   
\end{equation}   
\noindent   
where $s$ is the standard deviation and $n -1$ is the number of 
degrees of freedom.

This test has the advantage of using a standard probability
distribution function, but special care must be taken when considering
the cases where 3 (or more) pulses have been measured within a burst,
let us say A, B, and C.  We can form three differences, but obviously
one of them will not be independent of the others.  An approximation
could be done adjusting the number of degrees of freedom, but then
equation~(\ref{chi2}) will not be strictly valid. Numerical
simulations convinced us that, at least in this case, this is not a
good approximation.  The results shown in Table~\ref{Ttests} from the
$\chi^{2}_{a}$ test were obtained selecting pairs (A, B) and (B, C).
We verified that our conclusions are not affected by
this choice.
The dispersions expected assuming independency are
$\sqrt{2} \sigma_{\gamma}=1.02$ and $\sqrt{2} \sigma_{\eta}=0.96$, and the
standard deviations obtained are $s_{\gamma}=0.39$ and $s_{\eta}=0.30$,
respectively.
In all cases, the $P$-values are very small for both indices
$\gamma$ and $\eta$. Therefore we can confidently reject $H_{\rm 0}$
in favor of the alternative hypothesis, i.e., the distribution of
PLHIC indices over a single burst is narrower than the one observed
over the whole sample of GRBs.

\begin{figure*}[h]
\epsscale{0.8}
\plotone{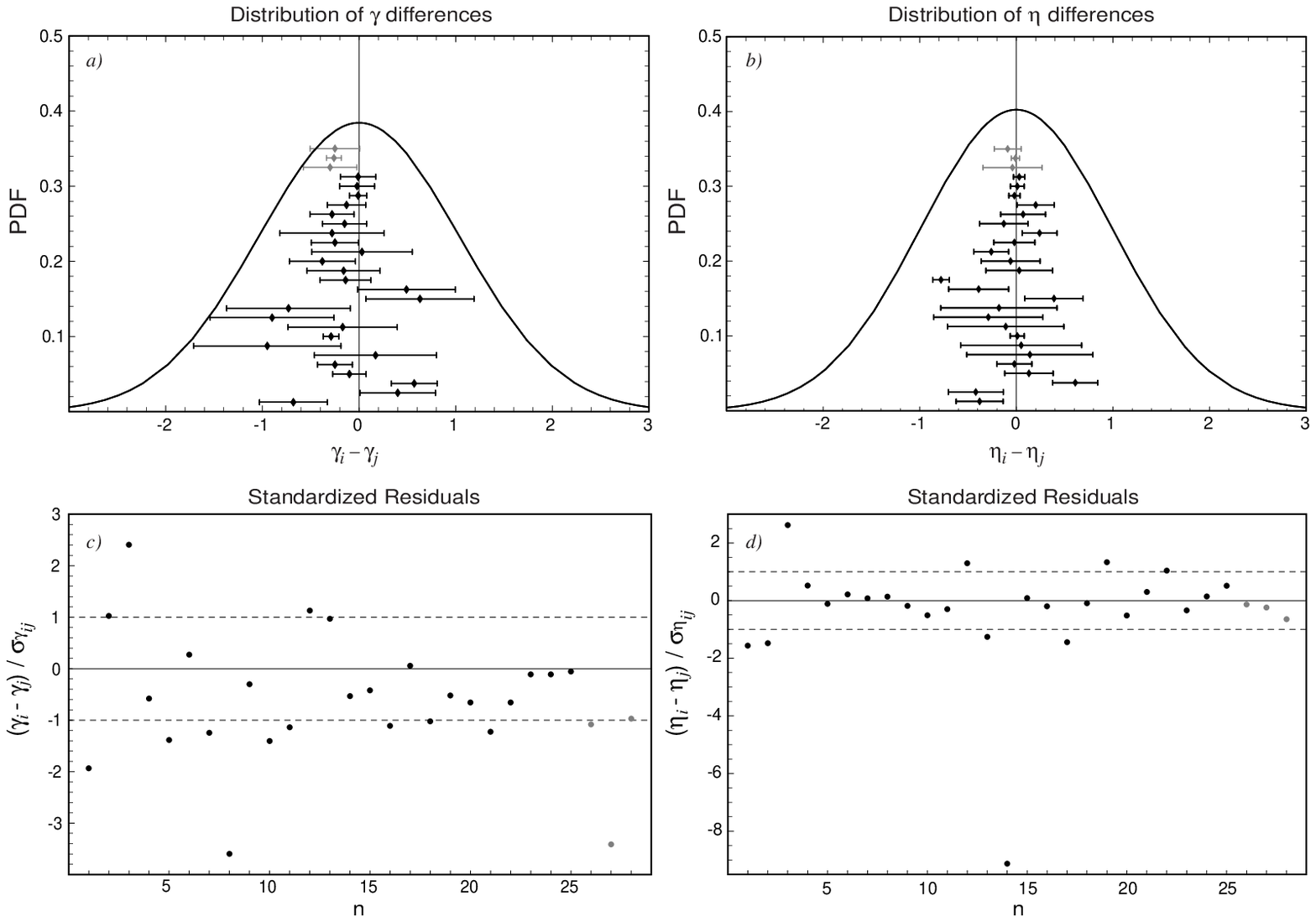}  
\caption{$a$), $b$) Probability density function (PDF) of the
differences, formed between the PLHIC indices $\gamma$ and $\eta$,
from pulses of different bursts (which are obviously independent)
follows approximately the normal distributions shown by the solid
curves.  Their dispersion are $\sqrt{2} \sigma_{\gamma}=1.02$ and
$\sqrt{2} \sigma_{\eta}=0.96$, respectively.  Differences between
indices within a single burst, e.g., $\eta_{i}-\eta_{j}$, should
follow the same distribution if they were independent too. The graph
shows instead that the 25 differences, taken from 35 pulses studied,
have a smaller dispersion (all differences are done in temporal
order). The standard deviations calculated are $s_{\gamma}=0.39$ and
$s_{\eta}=0.30$. Note that the dispersion and the error bars of the
$\eta$ differences are slightly smaller than those of
$\gamma$. Furthermore, most error bars cover the origin, implying
equal values to within the uncertainties. The ordinate position of the
points is arbitrary, but the order of the studied bursts is temporal
(see Table~\ref{Tmultiple}). The extra points in grey represent cases
belonging to the {\it track jump} sample (Table~\ref{Tjumps}).  $c$),
$d$) Standardized residuals are shown. Assuming that the indices are
constant within each burst, the observed spread should be entirely due
to the measurement uncertainties, presumably normally distributed. 
Dashed lines indicate the $\pm 1\sigma$
region. Note the ``outlier'' at number 14, related to the second pulse
of GRB 950624 (labeled 3648b in Table
\ref{Tsingle}). Note also that $\gamma$ differences tend to be 
negative, i.e., later pulses tend to have higher indices. This is 
likely to be an effect of the limited energy window. See the text for 
discussion.} 
\label{diff}  
\end{figure*}

In Figures~\ref{diff}$a$ and $b$ we illustrate how all possible index
differences, taking pulses in temporal order, compare with the normal
distribution assumed in $H_{\rm 0}$.  We had to run a numerical
simulation to find the non-standard probability distribution for the
variance of all index differences. We define the test statistic
\begin{equation}   
T_{a} \equiv \sum_{i=1}^{N}\;\; \sum_{j=1}^{M_{i}-1}\sum_{k=j+1}^{M_{i}}   
(X_{ij}-X_{ik})^2 ,   
\label{Ta}   
\end{equation}   
\noindent 
where $N$ is the number of bursts in the sample and $M_{i}$ 
is the number of measured pulses within burst~$i$.  This test is 
obviously independent of the order in which the differences are taken.  
Using a Monte Carlo method, we found the probability density function 
of $T_{a}$ assuming, as before, that each index $X_{ij}$ follows 
the general distributions found in \S~\ref{singlePL}, and using the 
normal approximation.  The results of this second test are presented 
in Table~\ref{Ttests}.  Again, $P$-values are very small for both 
indices, confirming the conclusions of the previous test.  The purpose 
of using first the $\chi_{a}^{2}$ test as an approximation is that, having a 
known analytical solution, it can estimate very low probabilities
without being limited by computational time.

We have now established that the distribution of indices among pulses
within a burst is narrower than the one between pulses of different
bursts.  Figures~\ref{diff}$a$ and $b$ show that most uncertainties in our
sample are much smaller than the dispersion of the general
distribution. Note also that most error bars cover the origin,
suggesting that individual index differences are not significant.  We
will now consider whether the index dispersion within a burst could be
attributed almost entirely to the stochastic uncertainties.
Figure~\ref{diff}$c$ and $d$ show the standardized residuals of all
possible differences $Y_{i}$ between pulses of the same burst in our
sample.  The proportion of data points within integer multiples of
$\sigma$ is consistent with a Gaussian distribution, except for an
``outlier'' in the $\eta$ set, at number 14. This point and point
number 12 are both differences taken against the second pulse of
GRB~950624 (pulse~3648b; see Table~\ref{Tsingle}) which is a fairly
weak but long pulse.  It precedes and slightly overlaps a strong pulse
(3648c), which could produce a systematic error in our measurements.
Note also, however, in Table~\ref{Tsingle} that comparing the
estimated index uncertainties almost always $2 \gtrsim
\Delta\gamma/\Delta\eta \gtrsim 1$. One exception is this case where
$\Delta\gamma/\Delta\eta \approx 8$. This is why for the $\gamma$
differences (Fig.~\ref{diff}c) the corresponding point is within
$\pm 1\sigma$.  The cause of the large deviation is most likely due to
a casual alignment of data points that resulted in an underestimation
of the $\Delta\eta$ uncertainty.  This fact stresses the importance of
having high selection requirements. If we had tried to expand our
sample to include even weaker or shorter pulses, with fewer time-bins,
it would have been at the cost of increasing the risk of systematic
errors.  Since these are difficult to identify and quantify, they can
easily distort the statistics.

Assuming that the indices are constant within a burst, we will
estimate the probability to exceed the observed dispersion. The uncertainties
will be assumed to be drawn from a normal distribution. Therefore the
index differences $Y \in \{X_{ij}-X_{ik}\}$ should follow 
$Y_{i} \sim  N(0,\sigma_{i})$, where each $\sigma_{i}^{2}$ is found by
adding the corresponding $X_{i}$ variances. We define, in a similar way as before,
a test statistic 
\begin{equation}   
T_{b}  \equiv \sum_{i=1}^{N}\;\;
\sum_{j=1}^{M_{i}-1}\sum_{k=j+1}^{M_{i}}  
\frac{(X_{ij}-X_{ik})^2}{\sigma^{2}_{ij}+\sigma^{2}_{ik}}  ,    
\label{Tb}   
\end{equation}   
\noindent   
where the summation considers all possible differences. Since not all
differences are independent, we have to solve it numerically. On the
other hand, if we restrict the sum to independent terms
\begin{equation}   
\chi^2_{b} \equiv \sum_{i=1}^{N}\;\;
\sum_{j=1}^{M_{i}-1}  
\frac{(X_{ij}-X_{ij+1})^2}{\sigma^{2}_{ij}+\sigma^{2}_{ij+1}}  ,    
\label{chi2b}   
\end{equation}   
\noindent  
the test becomes a standard $\chi^2$ statistic.

The results of these last two tests are summarized in the second part of Table
\ref{Ttests}.  For the $\gamma$ indices we found relatively low $P$-values
for both tests, but we consider that, at the 0.07 probability given by the
more precise test ($T_b$),  the hypothesis of invariance of
the $\gamma$ index is {\it marginally acceptable}. The results concerning the $\eta$
indices depend on the inclusion or rejection of  pulse number 3648b
mentioned above.  The {\it outlier} term completely dominates the sums in
equations~(\ref{chi2b}) and (\ref{Tb}).  Its inclusion gives
$P$-values $\ll 10^{-5}$ for both tests. Obviously, a $9\sigma$
difference can not be explained in terms of Gaussian uncertainties. On the
other hand, rejecting this data point, we obtained high $P$-values,
implying that the observed differences between $\eta$ indices are most 
likely due to the measurement uncertainties.

Finally, considering again Figure~\ref{diff}, it is apparent 
that while the $\eta$
differences have approximately a zero mean, the $\gamma$ differences
tend to be negative, i.e., later pulses have generally larger indices.
Taking the averages of the standardized residuals, the means are $-0.037$
and $-0.49$ for the $\eta$ and $\gamma$ differences, respectively.  We
calculated, in the $\gamma$ case, whether the non-zero mean deviation is
significant.  For that purpose a similar statistic to
equation~(\ref{Tb}) was used, but without squaring the terms.  We found a
0.03 probability (two-sided level) to exceed the observed value
assuming a zero mean, indicating a significant deviation.

\section{Discussion}\label{discussion}

\subsection{Comparison between the $F$ and $\varphi$ methods} \label{phidiscussion}

The introduction of the $\varphi$-method for HIC studies was motivated in
\S~\ref{phimethod}. It was shown (Fig.~\ref{gammaeta}) that there
is a good average correspondence between the indices $\gamma$ and
$\eta$. The scatter around an exact correspondence was to be
expected. As mentioned above, additional flux components can
be contributing to the estimated flux affecting the resulting
PLHIC index.  The {$\varphi$}-method is less dependent on this. 
Furthermore, the exact equivalence between the power law indices is
valid  only in the case that $\alpha$ and $\beta$ do not
vary with $E_{\rm pk}$, which they do in many bursts. 
Furthermore, the flux $F$ is found by integrating the deconvolved
spectrum and the deconvolution is model-dependent.  This could
introduce additional scatter into the PLHIC relation.

We have used both $F$ and $\varphi$ in our studies and the 
results lead basically to the same general conclusions, albeit
some important differences should be noted. 
In our analysis of the index distribution within a single burst
(Fig.~\ref{diff}), the $\eta_{i}-\eta_{j}$ differences show a narrower
distribution than the corresponding $\gamma_{i}-\gamma_{j}$.  In
\S~\ref{Fvsphi} we concluded that the new approach gives better
correlations. This implies that, in general, the estimated
uncertainties of $\eta$ are smaller. Even so, when testing the null
hypothesis of index invariance using equation (\ref{Tb}), $H_0$ is
marginally accepted in the $\gamma$ case (probability $p=7\%$) , while
it is fully consistent for $\eta$ ($p=64\%$).  We have also shown that
the general distributions of $\eta$ and $\gamma$ (Fig.~\ref{histo})
are very similar. Their means and dispersions are equal to within the
uncertainties, thus covering about the same range of values. All these
facts strongly suggest that the underlying distribution within a
single burst is indeed very narrow, that the observed larger spread
can be attributed to the uncertainties of the measurements, and that a
better estimator of the HIC will consequently reduce this
spread. Obviously, this is also valid for the determination of the
general distribution, but in this case the observed reduction is
comparable to the standard deviation uncertainties.

We also discussed in \S~\ref{tests} that within a burst, later pulses
seem to have larger $\gamma$ indices, although the difference is not
highly significant. This trend is not observed in the $\eta$
values. This might be a systematic error overcome by the
$\varphi$-method. Later pulses are usually softer, and therefore
closer to the lower energy limit of observation.
See, for example, the burst trigger~2083 in Figure~\ref{2083}. The
greater slope of the second pulse PLHIC, quantified in  panel ($c$), is
of the same order as  the one obtained in comparative simulations of
the methods, assuming a constant spectral shape (i.e., $\alpha$ and
$\beta$ fixed in eq.~[\ref{band}]). A larger sample is needed to
verify the trend, but this suggests that indeed the new method is less 
affected by the detector energy window.

The $\varphi$-method thus provides a better way of studying detailed
features in the observed HICs. Finally, the better invariance within a
burst shown by the $\eta$ indices may allow us to recognize hidden pulses
in the light curve of the burst, as will be discussed next.

\subsection{Track Jumps and Overlapping Pulses} \label{jumpdiscussion}

\begin{figure}[ht!] 
\epsscale{0.8}
\plotone{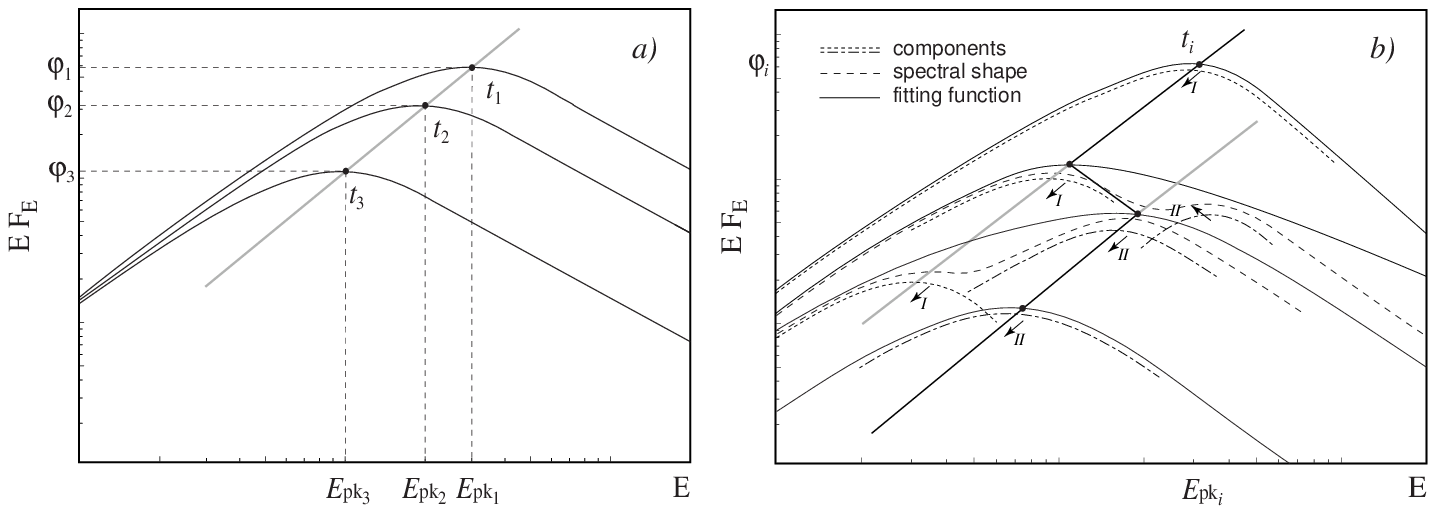} 
\caption{$a$) Schematic spectral evolution in the  $EF_{\rm E}$
representation.
During  pulse decays, the spectral peaks evolve in time following a power 
law. 
$b$) Schematic two-component model explaining {\it track jumps}.
Each component evolves in a similar way, following 
approximately parallel tracks when decaying.  However, the second is 
delayed in time and the ``jump'' occurs when it overcomes the first 
component. See the text for further details.} 
\label{model} 
\end{figure}

Figure~\ref{model}$a$ illustrates the observed evolution of
$\varphi(E_{\rm pk})$ along a {\it track} during the decay phase of a
single pulse. 
When many pulses are present in the light curve of a burst, and even
though there is no apparent general trend during the rise phases, we
have found that they follow almost parallel tracks during the decay
phases. To observe this behavior clearly, it is necessary that the
pulses do not overlap significantly. Also, they must be bright and
long enough to establish the direction of the {\it track}, see section
\ref{selection}.

In section \ref{jumps}, we presented the case GRB~950104
(trigger~3345) with marginally separated pulses, which exhibits a {\it
track jump} in its HIC. We consider this as a limiting case, where it
can still be claimed, from the analysis of the time history alone,
that a secondary pulse is present (see Fig.~\ref{3345&6397}a).  This
fact, together with the slope agreement, within the uncertainties,
between the first pulse track and the tracks in the second pulse,
makes it an interesting example among the {\it track jump}
cases.
   
Analogous HIC behaviors were found in a further three cases, shown in 
section \ref{jumpsII}, but there no visible secondary pulse was seen.  
For instance, GRB~970925 (trigger~6397; see Fig.~\ref{3345&6397}$b$) 
shows an apparent change in the overall decay, which could be due to a 
hidden pulse. However, there is an implicit assumption of the pulse
shape being an exponential. Nevertheless, in the present study we do
not want to make such assumptions concerning the shape of the pulses.

A simple interpretation of the {\it track jump} is to assume that it
is produced by an overlapping secondary pulse. From this point of
view, this feature is present also in cases like GRB 970420 (trigger
6198) shown in Figure~\ref{6198}.  The only difference is that there
we can resolve the individual pulses in the light curve.  Due to the
strong temporal evolution of $\Epk$ observed in all presumed single
pulses, two overlapping pulses will show, if all pulses within a burst
have similar spectral shapes and behaviors, a continuous change in the
way the individual spectral shapes overlap each other.  These {\it
fundamental} shapes may be fairly constant, but the combined evolution
in the $E\FE$--$E$ plane will give the impression of an overall
change.  In Figure~\ref{model}$b$, we illustrate such a model of two
spectral components. The components follow identical {\it tracks}
during their decay phases.  As the peak energy of the main pulse
declines, a harder component appears at the rise phase of the
secondary pulse.  At some point this component starts to dominate the
total spectrum and the fitting routine finds a better $\chi^{2}$
value, shifting the Band~et~al.\ function parameters to peak at a
higher $\Epk$, and the jump occurs.  Note that due to the broad shape
of the spectrum, the changes in the values of the asymptotic slopes of
the Band~et~al.\ function are an artifact of the fitting.

The uncertainties of the parameter $E_{\rm pk}$ are estimated by
evaluating the confidence region around the $\chi^2$ minima of
the fits.  When the spectral shape has a local
maximum close to the absolute maximum, an overestimation of the
$E_{\rm pk}$ uncertainties occurs.  These uncertainties are of the same
order as the energy interval that separates the tracks.  But the very
good alignment of the data points along the tracks indicates that the
shape features are real and not a random occurrence.

In studies of the spectral/temporal evolution, other workers (e.g.,
\cite{Kar95}; \cite{LK96}) have fixed the parameters $\alpha$ and $\beta$ 
to reduce the estimated uncertainties of $E_{\rm pk}$.  For instance,
the parameter values of the time-integrated spectrum can be used.
Such a solution provides more ``stable'' $E_{\rm pk}$ values, but a
case like GRB~950104 (trigger~3345) would be missed.  It smoothes out
the {\it track jump} in such a way that a single, steeper track is
obtained for the whole decay. Also a coarser time binning (see 
Ryde \& Svensson 2000c) makes the jump ``invisible''.

Based on this discussion we propose a pulse identification method for
cases of extensive overlap,
under the following assumptions.  First, the GRB pulses are assumed
to actually follow a PLHIC during their decays.  We may not observe this
because the pulse is too weak or short, because the observation energy window
is inadequate, or because there are unresolved secondary pulses present.
Second, we have to assume that the correlation index is
approximately constant within a burst. 
We have shown four cases where the pulse separation is evident in the
hardness-intensity plane but not as clearly in the light curve.
This pulse separation is not so easily accomplished by traditional methods
(see \S \ref{selection}),
but can be used as an auxiliary tool. However, to identify {\it track
jumps} with some confidence, the distance between tracks should be
larger than the observed dispersion along individual tracks. Also a
minimum number of points on each track is needed to produce a
reasonable fit.  

We found few cases in our BATSE sample that meet these
requirements. However, future instrumentation will allow higher
time-spectral resolution and these criteria may then have a wider
application.

\section{Constraints on  GRB Models} \label{models}   
   
The conditions that determine a particular PLHIC index may depend, in
principle, on many physical factors of the emitting system, e.g., the
density, composition, radiation processes, geometry,
etc. This fact is reflected by the broad distribution of the indices
$\eta$ and $\gamma$ that we found when studying pulses from different
bursts (see Fig.~\ref{histo}).  The internal models (see, e.g.,
\cite{kob97}, \cite{dai98}) rely on the assumption of a random 
distribution of some shell parameters, such as the density, the
kinetic energy or the momentum, to explain the chaotic time histories
observed but also to obtain high efficiencies. The sequence of
collisions and merging between different shells follows a stochastic
process that will hardly consistently reproduce identical physical
conditions in each emitting shock front.  On the contrary, most
probably it will statistically reflect almost the same diversity of
conditions found when we observe many bursts. Based on these models,
one would expect a similar parameter dispersion when pulses are taken
from the same or different bursts.  On the other hand, it is
conceivable that in the case bursts were very anisotropic, differences
due to the viewing angle would be enhanced by the relativistic beaming
and this could become a dominating factor.  In spite of that, the
internal models will still need to produce a narrow PLHIC index
dispersion within a burst.
The approximate constancy of the PLHIC index is more easily explained
in terms of a single system where an active region is regenerated or
different parts become active at different times, as it is assumed in
many variations of the external shock scenario (see, e.g., \cite{fen99}).

\section{Conclusions} \label{conclusions}  
  
We have introduced a new method to study the hardness-intensity
correlation in GRBs. Instead of measuring the energy flux, we use the
value of $\varphi = E F_{\rm E}$ at $E=E_{\rm pk}$, which differs by
an approximately constant factor. This method has many
advantages. First of all, the correlations become stronger. Second,
the $\varphi$-method is less dependent on the spectral window of
the observations.  Third, it disregards the effects of soft components
and additional weak pulses with spectra which overlap the one being
studied. In cases where the spectrum consists of two approximately
equally strong components (from different pulses) the $\varphi$-method
captures the evolution of one of the components and disregards the
second. These cases produce broad spectra which are not very well
modeled by the Band et al.\ function. Fourth, the resulting flux
measure, $\varphi$, is less sensitive to the details of the
deconvolution, e.g., the uncertainties in the spectral shape.

We constructed a sample of prominent GRB pulses consisting of 82
cases. Of these, we found that at least $75 \%$ exhibited moderately
good (or better) PLHICs.  Among the poorly correlated cases, we could
recognize some technical problems that were affecting our
analysis. Thus, the applicability of the PLHIC model may be even
larger. 

A specific feature was found in the hardness-intensity evolution of
some seemingly single pulses. In the HIC diagram of these cases, there
is a sharp transition from one power law track to a second, both with
the same PLHIC index.  This feature is denoted as a {\it track jump}
and in half of the cases, it coincides with a weak feature in the light
curve.  This could be explained as being produced by overlapping
pulses.  The {\it track jumps} in the HIC diagram reveal the
transition to a new pulse and it can be used as an aid for pulse
identification.

We established that the distribution of the PLHIC index is 
narrower for pulses within a burst than the distribution for pulses 
from different bursts. The latter can be approximated by a normal 
distribution. The dispersion within a burst could be entirely attributed  
to the stochastic uncertainties and the index is thus practically invariant. 

These results demand a physical model to be able to reproduce multiple
pulses from an individual burst with similar characteristics.  At the
same time, however, pulses from an ensemble of different bursts should
exhibit a diversity that is larger.  This is particularly relevant
when comparing the external versus the internal models. The latter
models require a large diversity of properties of the colliding shells
giving rise to the pulses, both within a burst and among bursts.

Changes in the observational conditions among different bursts, such
as viewing angle, could possibly account for  the
disparity between the index distributions. These matters will be
considered in future research.

\acknowledgments  

The present investigation made use of data obtained through the HEASARC Online
Service provided by NASA's Goddard Space Flight Center.  We are
grateful to the GRO Science Support Center for support. We would like
to thank Roland Svensson for enlightening and interesting discussions
as well as detailed comments on the manuscript,
and Stefan Larsson, Garrelt Mellema, and Karin Ryde for
various help. We also appreciate the valuable comments
given by Andrei Beloborodov and Juri Poutanen.
We acknowledge financial support from the Anna-Greta and
Holger Crafoord Fund, the Gustaf and Ellen Kobb's Stipend Fund at
Stockholm University, the Swedish Natural Science Research Council
(NFR), and the Swedish National Space Board.  FR wishes to express his
gratitude to Robert Preece, Jerry Fishman and Bill Paciesas for their
hospitality during his visit to NASA's Marshall Space Flight Center.

 
\newpage 

\newpage 
   
 
\begin{deluxetable}{lcccc} 
\scriptsize
\tablewidth{8.3cm} 
\tablecaption{Sample of 82 pulses in 66 GRBs} 
\tablehead{ 
\colhead{Burst} &\colhead{ trigger} &\colhead{LAD} 
 &\colhead{ $t_{\rm max} [ s ]$} &\colhead{$n_{\rm bins}$} }
\startdata 
GRB910627 &451	&  4	& 5.6     	& 5 	    \nl 
GRB910897 &647	&  0	& 3.2/14.0 	& 8/8 	    \nl 
GRB910814 &676	&  2	& 51.5 		& 5 	    \nl 
GRB910927 &829	&  4	& 6.3 		& 8  \tablenotemark{TJ} \nl 
GRB911016 &907 	&  1	& 1.4 		& 14 	   \nl 
GRB911031 &973 	&  3	& 2.7/24.2	& 19/12     \nl 
GRB911104 &999 	&  2	& 4.0 		&  5	    \nl 
GRB911118 &1085 &  4	& 8.7 		& 14 	    \nl 
GRB911126 &1121 &  4	& 21.9/24.7	& 8/6       \nl 
GRB911202 &1141	&  7	& 4.0/9.7 	& 8/8 	    \nl 
GRB920221 &1425	&  7	& 5.9 		&  5	    \nl 
GRB920502 &1578	&  5	& 8.0 		&  9	    \nl 
GRB920525 &1625	&  4	& 5.0/12.6 	& 12/5 	    \nl 
GRB920623 &1663	&  4	& 17.1 		&  10	    \nl 
GRB920718 &1709	&  7	& 2.1 		&  6	    \nl 
GRB920830 &1883 &  0	& 0.8 		&  7	    \nl 
GRB920902 &1886	&  5	& 8.3 		&  10        \nl 
GRB921003 &1974	&  2	& 1.3   	& 5 	    \nl   
GRB921123 &2067	&  1	& 21.2 		&  8	    \nl 
GRB921207 &2083	&  0	& 1.1/8.6 	& 22/9 	    \nl 
GRB930201 &2156	&  1	& 14.9 		& 12	      \nl 
GRB930425 &2316	&  1	&  14.4 	& 12 	    \nl 
GRB930612 &2387	&  6	& 5.9 		&  19	    \nl 
GRB931221 &2700	&  3	& 53.4 		&  8	    \nl 
GRB940329 &2897	&  4	& 8.2/24.0 	& 5/5       \nl 
GRB940410 &2919	&  6	& 0.5 		& 13	    \nl 
GRB940429 &2953	&  3	& 5.7/13.0 	& 9/7  	    \nl 
GRB940623 &3042	&  1	& 7.8 		& 14	    \nl 
GRB940708 &3067	&  6	& 2.4 		& 19	    \nl  
GRB941026 &3257	&  0	&  4.5		& 13 	    \nl 
GRB941121 &3290	&  0	& 39.7 		& 10 	    \nl 
GRB950104 &3345	&  1	& 5.8  		& 7 \tablenotemark{TJ}  \nl 
GRB950325 &3480	&  3	& 0.2 		& 10 	    \nl 
GRB950403 &3491	&  3	& 7.7 		&  16	    \nl 
GRB950403 &3492	&  5	& 5.3 		&  15	    \nl 
GRB950624 &3648	&  3	& 2.7/22.7/40.9	& 6/5/10    \nl 
GRB950818 &3765	&  1	&  66.1		&  12	    \nl 
GRB950909 &3788	&  3	&  27.2		&  14	    \nl 
GRB951016 &3870	&  5	&  0.5		&  8	    \nl 
GRB951102 &3891	&  2	&  33.3		&  7	    \nl 
GRB951213 &3954	&  2	&  0.8		&  16	    \nl 
GRB960113 &4350	&  1	&   13.8	&  6       \nl 
GRB960124 &4556	&  5	&  1.8/3.6	&  6/5      \nl 
GRB960530 &5478	&  2	&  1.9		&  7	    \nl 
GRB960531 &5479	&  0	& 52.5		&  5	    \nl 
GRB960708 &5534	&  5	& 1.7 		&  5	    \nl 
GRB960804 &5563	&  4	& 1.4 		&  6	    \nl 
GRB960807 &5567	&  0	& 11.8 		&  10	    \nl 
GRB960912 &5601	&  0	& 1.9   	& 10  \tablenotemark{TJ}  \nl 
GRB961001 &5621 &  2	& 3.9/7.2 	&  6/5 	    \nl 
GRB961009 &5628	&  1	& 10.2 		&  5	    \nl 
GRB961102 &5654	&  5	& 19.0 		&  14	    \nl 
GRB961126 &5697	&  6	& 0.6		&  6	    \nl 
GRB970111 &5773	&  0	& 8.1/17.3/19.4 & 8/5/6     \nl 
GRB970223 &6100	&  6	& 8.2 		& 16 	    \nl 
GRB970420 &6198	&  4	& 4.2/6.1/9.9 	& 5/8/5     \nl 
GRB970815 &6335	&  7	& 0.8	 	& 7 	    \nl 
GRB970925 &6397	&  7	& 2.8 	 	& 12 \tablenotemark{TJ} \nl 
GRB971127 &6504	&  2	& 3.2 		& 5  	    \nl 
GRB980125 &6581	&  0	& 47.6 		& 7 	    \nl 
GRB980301 &6621	&  1	& 32.7 		& 6 	    \nl 
GRB980306 &6629	&  1	& 215 		& 6 	  \nl 
GRB980306 &6630 &  3	& 2.1 		& 9 	    \nl 
GRB980821 &7012	&  0	& 2.9 		& 6 	    \nl 
GRB990102 &7293	&  6	& 3.3 		& 10 	    \nl 
GRB990123 &7343 &  0	& 37.7 		& 8 	    
\enddata 
\tablenotetext{TJ}{ Case with a {\it track jump} feature in its 
hardness-intensity temporal evolution}
\label{Tsample} 
\end{deluxetable} 
  
\newpage 
 
 
\begin{deluxetable}{lccccllc} 
\scriptsize
\tablewidth{15cm} 
\tableheadfrac{.18}
\tablecaption{The HIC power law indices for 78 pulses in 62 GRBs} 
\tablehead{ 
\colhead{Pulse} 
& \colhead{$\eta$} & \colhead{$\gamma$ } & \colhead{$R^{2}_{\eta}$} & \colhead{$R^{2}_{\gamma}$} &
\colhead{ $P_{N}(R^{2}_{\eta})$} & \colhead{$\Delta\eta/\eta$} &
\colhead{Comments} }
\startdata 
451b   & $  1.36  \pm  0.24   $ &   $  1.38  \pm  0.33   $ &  0.9166  & 0.8509  &  $1.0\,\times {{10}^{-2}}$  &  0.18  & \nl 
647a   & $  2.12  \pm  0.22   $ &   $  1.80  \pm  0.25   $ &  0.9382  & 0.8868  & $ 7.6\,\times {{10}^{-5}}$  &  0.10   &$A$ \nl 
647b   & $  2.54  \pm  0.18   $ &   $  1.40  \pm  0.30   $ &  0.9692  & 0.7888  & $9.2\,\times {{10}^{-6}}$   &  0.071  & $A,B$\nl 
676    & $  3.30  \pm  0.40   $ &   $  2.80  \pm  0.25   $ &  0.9561  & 0.9788  &  $4.0\,\times {{10}^{-3}}$   &  0.12   & $A,B$\nl 
907    & $  2.34  \pm  0.15   $ &   $  2.54  \pm  0.20   $ &  0.9512  & 0.9422  & $3.1\,\times {{10}^{-9}}$   &  0.064  &$A,B$ \nl 
973a   & $  1.45  \pm  0.21   $ &   $  1.28  \pm  0.21   $ &  0.7371  & 0.6821  & $2.6\,\times {{10}^{-6}}$   &  0.14   &$A,B$  \nl 
973b   & $  0.84  \pm  0.10   $ &   $  0.71  \pm  0.11   $ &  0.8590  & 0.8041  &  $1.5\,\times {{10}^{-5}}$   &  0.12   &$A$ \nl 
999    & $  1.90  \pm  0.10   $ &   $  1.60  \pm  0.10   $ &  0.9905  & 0.9825  & $3.9\,\times {{10}^{-4}}$   &  0.053  &$A,B$ \nl 
1085   & $  1.79  \pm  0.025  $ &   $  2.05  \pm  0.04   $ &  0.9978  & 0.9954  & $ < 10^{-10}$               &  0.014  &$A,B$ \nl 
1121a  & $  1.30  \pm  0.15   $ &   $  1.00  \pm  0.10   $ &  0.9318  & 0.9172  &  $1.0\,\times {{10}^{-4}}$   &  0.12   &$A,B$ \nl 
1121b  & $  1.17  \pm  0.20   $ &   $  1.10  \pm  0.14   $ &  0.9150  & 0.9385  &  $2.8\,\times {{10}^{-3}}$  &  0.17   & \nl 
1141a  & $  0.84  \pm  0.10   $ &   $  0.51  \pm  0.10   $ &  0.9198  & 0.8117  &  $1.7\,\times {{10}^{-4}}$  &  0.12   &$A,B$ \nl 
1141b  & $  0.86  \pm  0.15   $ &   $  0.76  \pm  0.15   $ &  0.8781  & 0.8400  &  $5.9\,\times {{10}^{-4}}$   &  0.17   & \nl 
1425   & $  2.70  \pm  0.50   $ &   $  2.55  \pm  0.50   $ &  0.8922  & 0.9060  &  $1.6\,\times {{10}^{-2}}$  &  0.19   & \nl 
1578   & $  2.80  \pm  0.40   $ &   $  2.95  \pm  0.55   $ &  0.8696  & 0.8038  &  $2.5\,\times {{10}^{-4}}$   &  0.14   &$A,B$ \nl 
1625a  & $  2.37  \pm  0.30   $ &   $  2.12  \pm  0.31   $ &  0.8584  & 0.8188  &  $1.5\,\times {{10}^{-5}}$  &  0.13   &$A,B$ \nl 
1625b  & $  2.23  \pm  0.58   $ &   $  1.95  \pm  0.55   $ &  0.8315  & 0.8057  &  $3.1\,\times {{10}^{-2}}$  &  0.26   & \nl 
1663   & $  1.45  \pm  0.15   $ &   $  1.00  \pm  0.15   $ &  0.9139  & 0.8342  &  $1.6\,\times {{10}^{-5}}$   &  0.10   &$A,B$ \nl 
1709   & $  2.30  \pm  0.15   $ &   $  2.32  \pm  0.08   $ &  0.9865  & 0.9946  &  $6.9\,\times {{10}^{-5}}$   &  0.065  &$A,B$ \nl 
1883   & $  2.39  \pm  0.29   $ &   $  2.51  \pm  0.40   $ &  0.9303  & 0.8783  &  $4.5\,\times {{10}^{-4}}$   &  0.12   &$A,B$ \nl 
1886   & $  3.80  \pm  0.45   $ &   $  3.60  \pm  0.45   $ &  0.8965  & 0.8786  &  $3.3\,\times {{10}^{-5}}$  &  0.12   & $A,B$\nl 
1974a  & $  2.20  \pm  0.30   $ &   $  2.10  \pm  0.30   $ &  0.9537  & 0.9495  &  $4.3\,\times {{10}^{-3}}$  &  0.14   &$A,B$ \nl 
2067   & $  2.60  \pm  0.15   $ &   $  2.60  \pm  0.10   $ &  0.9852  & 0.9902  & $ 1.0\,\times {{10}^{-6}}$    &  0.058  &$A,B$ \nl 
2083a  & $  1.78  \pm  0.04   $ &   $  1.74  \pm  0.04   $ &  0.9889  & 0.9890  &  $ < 10^{-10}$       &  0.022  &$A,B$ \nl 
2083b  & $  1.77  \pm  0.06   $ &   $  2.03  \pm  0.07   $ &  0.9915  & 0.9926  & $ 1.7\,\times {{10}^{-8}}$   &  0.034  & $A$ \nl 
2156   & $  1.55  \pm  0.15   $ &   $  1.00  \pm  0.13   $ &  0.9117  & 0.8460  & $ 1.4\,\times {{10}^{-6}}$   &  0.097  &$A,B$ \nl 
2316   & $  0.64  \pm  0.23   $ &   $  0.60  \pm  0.30   $ &  0.4395  & 0.2870  &  $1.8\,\times {{10}^{-2}}$   &   0.35    & \nl 
2387   & $  1.09 \pm   0.17   $ &   $  0.83  \pm  0.33   $ &  0.7035  & 0.2695  &  $7.2\,\times 10^{-6}$     &   0.16    & \nl 
2700   & $  1.70  \pm  0.45   $ &   $  1.70  \pm  0.45   $ &  0.7067  & 0.6867  &  $8.9 \,\times {{10}^{-3}}$  &  0.26   & \nl 
2897b  & $  1.24  \pm  0.45   $ &   $  1.47  \pm  0.40   $ &  0.7344  & 0.7886  &  $6.4\,\times {{10}^{-2}}$  &  0.36   & \nl 
2897c  & $  1.42  \pm  0.40   $ &   $  2.20  \pm  0.50   $ &  0.8177  & 0.8179  &  $3.5\,\times {{10}^{-2}}$  &  0.28   & \nl 
2919   & $  1.39  \pm  0.38   $ &   $  2.20  \pm  0.50   $ &  0.5478  & 0.3954  &  $3.8\,\times {{10}^{-3}}$  &  0.27   & \nl 
2953a  & $  1.61  \pm  0.30   $ &   $  1.28  \pm  0.27   $ &  0.8094  & 0.7577  &  $9.5\,\times {{10}^{-4}}$  &  0.19   & \nl 
2953b  & $  1.46  \pm  0.45   $ &   $  1.24  \pm  0.41   $ &  0.6785  & 0.5777  &  $2.3\,\times {{10}^{-2}}$  &  0.31   & \nl 
3042   & $  1.38  \pm  0.88   $ &   $  1.07  \pm  0.90   $ &  0.3809  & 0.7875  &  $1.9\,\times {{10}^{-1}}$  &  0.64   & \nl 
3067   & $  1.35  \pm  0.10  $ &   $   1.47  \pm  0.13   $ &  0.8963  & 0.8820  & $ 8.6\,\times {{10}^{-10}}$  &  0.074  &$A,B$ \nl 
3257   & $  2.80  \pm  0.20   $ &   $  2.60  \pm  0.20   $ &  0.9573  & 0.9268  & $ 7.0\,\times {{10}^{-9}}$    &  0.071  &$A,B$ \nl 
3290   & $  1.12  \pm  0.08   $ &   $  1.41  \pm  0.10   $ &  0.9625  & 0.9560  & $ 5.5\,\times {{10}^{-7}}$   &  0.071  &$A,B$ \nl 
3480   & $  2.07  \pm  0.15   $ &   $  1.90  \pm  0.13   $ &  0.9686  & 0.9653  & $ 2.7\,\times {{10}^{-7}}$   &  0.072  & $A,B$\nl 
3491   & $  1.75  \pm  0.09   $ &   $  1.59  \pm  0.10   $ &  0.9600  & 0.9477  &  $ < 10^{-10}$        &  0.051  &$A,B$ \nl 
3492   & $  1.65  \pm  0.10   $ &   $  1.50  \pm  0.10   $ &  0.9378  & 0.9272  & $ 3.2\,\times {{10}^{-9}}$   &  0.061  & $A,B$\nl 
3648a  & $  1.03  \pm  0.30   $ &   $  1.87  \pm  0.50   $ &  0.7705  & 0.7647  &  $2.1\,\times {{10}^{-2}}$   &  0.29   & \nl 
3648b  & $  0.64  \pm  0.03   $ &   $  1.24  \pm  0.25   $ &  0.9930  & 0.9027  &  $2.5\,\times {{10}^{-4}}$   &  0.047  &$A$  \nl 
3648c  & $  1.42  \pm  0.08   $ &   $  1.38  \pm  0.08   $ &  0.9743  & 0.9749  & $ 1.2\,\times {{10}^{-7}}$   &  0.056  & $A,B$\nl 
3765   & $  2.40  \pm  0.20   $ &   $  2.45  \pm  0.20   $ &  0.9438  & 0.9217  & $ 1.4\,\times {{10}^{-7}}$   &  0.083  &$A,B$ \nl 
3788   & $  1.67  \pm  0.26   $ &   $  1.35  \pm  0.35   $ &  0.7691  & 0.5571  &  $3.8\,\times 10^{-5}$     &  0.16    & \nl 
3870   & $  0.78  \pm  0.15   $ &   $  0.76  \pm  0.20   $ &  0.7927  & 0.7147  &  $3.0\,\times {{10}^{-3}}$   &  0.19   & \nl 
3891   & $  1.80  \pm  0.18   $ &   $  1.58  \pm  0.25   $ &  0.9498  & 0.8922  &  $2.0\,\times {{10}^{-4}}$   &  0.10   & $A,B$\nl 
3954   & $  1.25 \pm  0.20    $ &   $  0.97  \pm  0.21   $ &  0.7759  & 0.5907  &  $6.6\,\times 10^{-6}$     &   0.16    & (1) \nl 
4350   & $  0.82 \pm  2.8     $ &   $  1.0   \pm  4.3    $ &  0.0207  & 0.0137  &  $7.9\,\times {{10}^{-1}}$  &   3.4     & \nl 
4556a  & $  1.50  \pm  0.30   $ &   $  1.20  \pm  0.30   $ &  0.8756  & 0.7937  &  $6.1\,\times {{10}^{-3}}$   &  0.20   & \nl 
4556b  & $  1.47  \pm  0.17   $ &   $  1.36  \pm  0.23   $ &  0.9621  & 0.9180  &  $3.2\,\times {{10}^{-3}}$   &  0.12   & $A,B$\nl 
5478   & $  1.87  \pm  0.16   $ &   $  1.05  \pm  0.20   $ &  0.9664  & 0.8835  &  $7.1\,\times {{10}^{-5}}$   &  0.086  &$A,B$ \nl 
5479   & $  1.06  \pm  0.26   $ &   $  1.03  \pm  0.30   $ &  0.8423  & 0.7785  &  $2.8\,\times {{10}^{-2}}$   &  0.25   & \nl 
5534   & $  0.55  \pm  0.14   $ &   $  0.96  \pm  0.17   $ &  0.8415  & 0.9138  &  $2.8\,\times {{10}^{-2}}$   &  0.25   & \nl 
5563   & $  2.35  \pm  0.30   $ &   $  2.40  \pm  0.40   $ &  0.9226  & 0.9042  &  $2.3\,\times {{10}^{-3}}$   &  0.13   & $A,B$\nl 
5567   & $  3.96  \pm  0.53   $ &   $  3.68  \pm  0.50   $ &  0.8750  & 0.8761  &  $7.0\,\times {{10}^{-5}}$   &  0.13   & $A,B$\nl 
5621a  & $  1.26  \pm  0.26   $ &   $  0.97  \pm  0.32   $ &  0.8521  & 0.6893  &  $8.6\,\times {{10}^{-3}}$  &  0.21   & \nl 
5621b  & $  1.32  \pm  0.16   $ &   $  1.35  \pm  0.12   $ &  0.9570  & 0.9773  &  $3.8\,\times {{10}^{-3}}$   &  0.12   & $A,B$\nl 
5628   & $  1.05  \pm  0.25   $ &   $  0.75  \pm  0.20   $ &  0.8503  & 0.7850  &  $2.6\,\times {{10}^{-2}}$   &  0.24   & \nl 
5654   & $  2.00  \pm  0.50   $ &   $  2.00  \pm  0.50   $ &  0.5549  & 0.5728  &  $2.2\,\times {{10}^{-3}}$   &  0.25   & \nl 
5697   & $  3.75  \pm  0.70   $ &   $  4.50  \pm  0.70   $ &  0.8804  & 0.9057  &  $5.6\,\times {{10}^{-3}}$   &  0.19   & \nl 
5773a  & $  1.56  \pm  0.15   $ &   $  1.35  \pm  0.14   $ &  0.9573  & 0.9397  &  $2.5\,\times {{10}^{-5}}$   &  0.096  & $A,B$\nl 
5773b  & $  1.82  \pm  0.10   $ &   $  1.32  \pm  0.50   $ &  0.9868  & 0.7122  &  $6.5\,\times {{10}^{-4}}$   &  0.055  & $A$ \nl 
5773c  & $  1.58  \pm  0.15   $ &   $  1.60  \pm  0.20   $ &  0.9614  & 0.9261  &  $5.7\,\times {{10}^{-4}}$   &  0.095  &$A$  \nl 
6100   & $  2.03  \pm  0.18   $ &   $  1.79  \pm  0.20   $ &  0.8992  & 0.8549  & $2.3\,\times {{10}^{-8}}$    &  0.089  & $A,B$\nl 
6198a  & $  2.25  \pm  0.20   $ &   $  2.01  \pm  0.18   $ &  0.9793  & 0.9769  &  $1.3\,\times {{10}^{-3}}$   &  0.089  &$A$ \nl 
6198b  & $  2.38  \pm  0.15   $ &   $  2.16  \pm  0.14   $ &  0.9784  & 0.9766  & $ 3.2\,\times {{10}^{-6}}$   &  0.063  &$A,B$ \nl 
6198c  & $  2.18  \pm  0.12   $ &   $  2.29  \pm  0.14   $ &  0.9908  & 0.9892  &  $3.8\,\times {{10}^{-4}}$   &  0.055  & $A$ \nl 
6335a  & $  2.35  \pm  0.20   $ &   $  2.70  \pm  0.30   $ &  0.9727  & 0.9431  &  $4.2\,\times {{10}^{-5}}$   &  0.085  &$A,B$  \nl 
6504   & $  1.96  \pm  0.36   $ &   $  2.38  \pm  0.70   $ &  0.9068  & 0.8117  &  $1.2\,\times {{10}^{-2}}$   &  0.18   & \nl 
6581   & $  1.62  \pm  0.20   $ &   $  1.40  \pm  0.20   $ &  0.9374  & 0.9327  &  $3.4\,\times {{10}^{-4}}$   &  0.12   &$A,B$ \nl 
6621   & $  2.0  \pm  1.0    $ &   $   2.30  \pm  1.2    $ &  0.4892  & 0.4738  &  $1.2\,\times {{10}^{-1}}$   &  0.50   & \nl 
6629   & $  1.20  \pm  0.9    $ &   $  0.95  \pm  0.85   $ &  0.3274  & 0.2319  &  $2.4\,\times {{10}^{-1}}$   &  0.75   & \nl 
6630   & $  1.49  \pm  0.04   $ &   $  1.58  \pm  0.05   $ &  0.9944  & 0.9935  &  $3.8\,\times {{10}^{-9}}$   &  0.027  & $A,B$\nl 
7012   & $  1.91  \pm  0.44   $ &   $  1.60  \pm  0.40   $ &  0.8231  & 0.7448  &  $1.2\,\times {{10}^{-2}}$   &   0.23    & \nl 
7293   & $  2.04  \pm  0.12   $ &   $  2.05  \pm  0.22   $ &  0.9731  & 0.9145  & $ 1.4\,\times {{10}^{-7}}$   &  0.059  & $A,B$\nl 
7343   & $  1.50  \pm  0.50   $ &   $  1.10  \pm  0.50   $ &  0.5905  & 0.4201  &  $2.5\,\times {{10}^{-2}}$   &  0.33   & 
\enddata 
\tablecomments{Pulses are denoted by the burst trigger number and,
when it is necessary, labeled alphabetically following temporal
order. The labeling is consistent with the additional pulses presented
in Table~\ref{Tmultiple}.  In the comments column we specify the pulses
belonging to the subset $A$ consisting of 47 pulses from 39 bursts,
and defined by the condition $\Delta\eta/\eta
\leq 0.15$.  The subset $B$ is also indicated, and it is constructed
as $A$, but taking only one pulse per burst (choosing the one with
lowest $P_{N}(R^{2}_{\eta})$), therefore it comprises 39 pulses from
39 bursts.  Note that the four {\it track jump} cases from
Table~\ref{Tsample} are excluded here.}
\tablenotetext{1}{ Rejected as $\alpha$ is not constrained } 
\label{Tsingle} 
\end{deluxetable} 

\newpage   
 

\begin{deluxetable}{lcccclll} 
\scriptsize  
\tablewidth{15cm}  
\tablecaption{Extended Sample of Multi-Pulse Bursts}  
\tablehead{  
\colhead{Pulse} 
& \colhead{$t_{\rm max}[s]$} 
& \colhead{$n_{\rm bins}$} 
& \colhead{$\eta$} 
& \colhead{$\gamma$} 
& \colhead{$R^{2}_{\eta}$} 
& \colhead{ $P_{N}(R^{2}_{\eta})$} 
& \colhead{$\Delta\eta/\eta$}    
}  
\startdata 
451a   & 0.6  & 4 &   $  0.98  \pm  0.04  $ &   $  0.70  \pm  0.12   $ &  0.9963  & $  1.9\,\times {{10}^{-3}}  $ &  0.041   \nl 
451b   & 5.6  & 5 &   $  1.36  \pm  0.24  $ &   $  1.38  \pm  0.33   $ &  0.9166  & $  1.0\,\times {{10}^{-2}}  $ &  0.18    \nl 
1974a  & 1.3  & 5 &   $  2.20  \pm  0.30  $ &   $  2.10  \pm  0.30   $ &  0.9537  & $  4.3\,\times {{10}^{-3}}  $ &  0.14    \nl 
1974b  & 6.5  & 4 &   $  2.15  \pm  0.55  $ &   $  3.05  \pm  0.70   $ &  0.8796  & $  6.2\,\times {{10}^{-2}}  $ &  0.26    \nl 
2897a  & 5.9  & 4 &   $  1.13  \pm  0.40  $ &   $  1.30  \pm  0.40   $ &  0.7062  & $  1.6\,\times {{10}^{-1}}  $ &  0.35    \nl 
2897b  & 8.2  & 5 &   $  1.24  \pm  0.45  $ &   $  1.47  \pm  0.40   $ &  0.7344  & $  6.4\,\times {{10}^{-2}}  $ &  0.36    \nl 
2897c  & 24.0 & 5 &   $  1.42  \pm  0.40  $ &   $  2.20  \pm  0.50   $ &  0.8177  & $  3.5\,\times {{10}^{-2}}  $ &  0.28    \nl 
3345a  & 3.5  & 4 &   $  0.77  \pm  0.05  $ &   $  0.60  \pm  0.06   $ &  0.9909  & $  4.6\,\times {{10}^{-3}}  $ &  0.065   \nl 
3345b  & 5.8  & 3 &   $  0.79  \pm  0.03  $ &   $  0.61  \pm  0.065  $ &  0.9985  & $  2.5\,\times {{10}^{-2}}  $ &  0.038   \nl 
3345c  & 6.5  & 4 &   $ 0.76   \pm  0.05  $ &   $  0.62  \pm  0.17   $ &  0.9892  & $  5.4\,\times {{10}^{-3}}  $ &  0.066   \nl
\enddata   
\tablecomments{The multi-pulse sample comprises 35 pulses from 15
bursts, and it is an extension of the main sample that includes also 4
time-bin pulses (not listed in Table~\ref{Tsample}).  All the bursts
studied in this set have at least one pulse having more than 4
time-bins that belong to the main sample. Here we present only the
bursts with additional pulses for conciseness.  The whole set is,
listed by burst trigger number (number of pulses): 451(2) , {\it
647(2), 973(2), 1121(2), 1141(2), 1625(2),} 1974(2), {\it 2083(2),}
2897(3), 3345(3), {\it 3648(3), 4556(2), 5621(2), 5773(3), 6198(3)}
(cases in {\it italic} appeared in Table~\ref{Tsingle}) .  Pulse
3345 in Table~\ref{Tsample} is listed as a {\it track jump} case, and
it has been separated into two short pulses ($b$ and $c$) with 3 and 4
time-bins, respectively (see the discussion in the text).}
\label{Tmultiple}  
\end{deluxetable}  
  
\newpage

  
\begin{deluxetable}{lcclll}
\scriptsize 
\tablewidth{8.8cm}  
\tablecaption{Pulses with Track Jumps}  
\tablehead{  
\colhead{Pulse} 
& \colhead{$\eta$} 
& \colhead{$\gamma$} 
& \colhead{$R^{2}_{\eta}$} 
& \colhead{ $P_{N}(R^{2}_{\eta})$} 
& \colhead{$\Delta\eta/\eta$}    
}  
\startdata 
829a   &    $  2.87  \pm  0.30   $ &   $  3.20  \pm  0.12  $ &  0.9816  & $  9.2\times {{10}^{-3}}   $ &  0.10    \nl 
829b   &    $  2.91  \pm  0.06   $ &   $  3.50  \pm  0.25  $ &  0.9995  & $  2.5\times {{10}^{-4}}  $ &  0.021   \nl 
5601a  &    $  0.67  \pm  0.013  $ &   $  0.48  \pm  0.07  $ &  0.9992  & $  4.0\times {{10}^{-4}}   $ &  0.019   \nl 
5601b  &    $  0.68  \pm  0.04   $ &   $  0.74  \pm  0.03  $ &  0.9966  & $  3.7\times {{10}^{-2}}    $ &  0.059   \nl 
6397a  &    $  0.77  \pm  0.13   $ &   $  0.24  \pm  0.25  $ &  0.9265  & $  8.7\times {{10}^{-3}}   $ &  0.17    \nl 
6397b  &    $  0.86  \pm  0.05   $ &   $  0.49  \pm  0.06  $ &  0.9889  & $  5.0\times {{10}^{-4}}   $ &  0.058   \nl 
\enddata   
\label{Tjumps}  
\end{deluxetable}  
  
\newpage 
 
  
\begin{deluxetable}{lcc} 
\scriptsize  
\tablewidth{6cm}  
\tablecaption{Hypothesis Tests - $P$-values}  
\tablehead{  
\colhead{Test}    
&  \colhead{$\eta$}   
& \colhead{$\gamma$}}  
\startdata  
$\chi^{2}_{a}$ (Eq.~[\ref{chi2}])  & $   10^{-7} $       & $ 5  \times 10^{-6}      $ \nl
$T_{a}       $ (Eq.~[\ref{Ta}])    & $ < 10^{-5}     $ & $4  \times 10^{-5} $ \nl
$\chi^{2}_{b}$ (Eq.~[\ref{chi2b}]) & $ 7  \times 10^{-14} \, \Vert \, 0.38 $ \tablenotemark{\dagger} & 0.03  \nl
$T_{b}       $ (Eq.~[\ref{Tb}])    & $ < 10^{-5}  \phn \, \Vert \, 0.64 $ \tablenotemark{\dagger} & 0.07 \nl
\enddata
\tablenotetext{\dagger}{Results with and without the {\it outlier} data point discussed in the text}
\label{Ttests}
\end{deluxetable}

\end{document}